\documentclass[prd,12pt,nofootinbib,showpacs,preprint]{revtex4}
\usepackage[T1]{fontenc}
\usepackage{amsmath,amssymb}
\usepackage{epsfig}
\usepackage{url}
\usepackage{graphicx}
\usepackage[usenames,dvipsnames]{color}
\usepackage{slashed}
\usepackage[colorlinks,citecolor=blue]{hyperref}
\usepackage{color}
\usepackage{cancel}

\def \met{\not \! E_T }
\def \me{\not \! E }

\def\bar {\overline}

\def\be {\begin{equation}}
\def\ee {\end{equation}}
\def\beq {\begin{equation}}
\def\eeq {\end{equation}}
\def\bea {\begin{eqnarray}}
\def\eea {\end{eqnarray}}

\def\ifb{{\rm fb}^{-1}}

\def\beq{\begin{equation}}
\def\eeq{\end{equation}}
\def\barr{\begin{array}}
\def\earr{\end{array}}

\begin{document}
\preprint{HRI-RECAPP-2017-011}
%%%%%%%%%%%%%%%%%%%%%%%%%%%%%%%%%%%%%%%%%%%%%%%%%%%%%%%
\title{Lepton flavor violating Higgs decay at $e^+  e^-$ colliders}
\author{Indrani Chakraborty}
\email{indranichakraborty@hri.res.in}
\affiliation{Regional Centre for Accelerator-based Particle Physics, 
Harish-Chandra Research Institute, HBNI, Chhatnag Road, Jhusi, Allahabad - 211019, India} 
\author{Subhadeep Mondal}
\email{subhadeepmondal@hri.res.in}
\affiliation{Regional Centre for Accelerator-based Particle Physics, 
Harish-Chandra Research Institute, HBNI, Chhatnag Road, Jhusi, Allahabad - 211019, India}
\author{Biswarup Mukhopadhyaya}
\email{biswarup@hri.res.in}
\affiliation{Regional Centre for Accelerator-based Particle Physics, 
Harish-Chandra Research Institute, HBNI, Chhatnag Road, Jhusi, Allahabad - 211019, India}
%%%%%%%%%%%%%%%%%%%%%%%%%%%%%%%%%%%%%%%%%%%%%%%%%%%%%%%
\begin{abstract}
%\begin{center}
%{\bf Abstract }
%\end{center}
%%%%%%%%%%%%%%%%%%%%%%%%%%%%%%%%%%%%%%%%%%%%%%%%%%%%%%%
We estimate the smallest branching ratio for the Higgs decay channel $h \rightarrow \mu \tau$, 
which can be probed at an $e^+e^-$ collider and compare it with the projected reach at the 
high-luminosity run of the LHC.   
Using a model-independent approach, Higgs 
production is considered in two separate cases. In the first case, $hWW$ and $hZZ$ 
couplings are allowed to be scaled by a factor allowed by the latest experimental limits on $hWW$ 
and $hZZ$ couplings. In the second case, we have introduced higher-dimensional 
effective operators for these interaction vertices. Keeping BR($h\to\mu\tau$) as a purely phenomenological 
quantity, we find that this branching ratio can be probed down to $\approx 2.69\times 10^{-3}$ and $\approx 5.83\times 10^{-4}$ respectively, at the 250 GeV and 1000 GeV run of an $e^+e^-$ collider.  
\end{abstract} 
%%%%%%%%%%%%%%%
\maketitle
%%%%%%%%%%%%%%%
%%%%%%%%%%%%%%%%%%%%%%%%%%%%%%%%%%%%%%%%%%%%%%%%%%%%%%%%%
\section{Introduction}
\label{sec:intro}
%%%%%%%%%%%%%%%%%%%%%%%%%%%%%%%%%%%%%%%%%%%%%%%%%%%%%%%%%
After the discovery of the scalar resonance around 125 GeV at the LHC \cite{Aad:2012tfa,Chatrchyan:2012xdj}, 
efforts are under way to determine whether it is indeed the Standard Model (SM) Higgs boson. The spin, 
parity and couplings \cite{Choi:2002jk,Corbett:2012dm,Ellis:2012hz,Stolarski:2012ps,Alves:2012fb,Ellis:2012jv,Ellis:2012mj,Banerjee:2012xc,Cacciapaglia:2012wb,Moreau:2012da} of this new member are found to be in good agreement so far with the SM expectation. The couplings between the Higgs boson and gauge 
bosons, though consistent with the predictions of the SM \cite{Aad:2014eha,Khachatryan:2014ira, 
Aad:2014eva,Chatrchyan:2013mxa,ATLAS:2014aga,Aad:2015ona,Chatrchyan:2013iaa},  still leave some scope 
for deviation, thus keeping alive the possibility that it is `a Higgs' rather than `the Higgs'. The former possibility keeps up the hope
of addressing the yet unanswered questions like finding a suitable dark matter candidate, non-zero neutrino masses and mixing and baryon asymmetry of the universe. 
Side by side, possible hints of new physics may still be hidden in the 
considerable amount of imprecision remaining in the measurement of couplings between Higgs and heavy fermion pairs like
$\tau^+ \tau^-$, $b \bar{b}$ \cite{Aad:2014xzb,Chatrchyan:2013zna,Aad:2015vsa,Chatrchyan:2014nva} and of course, the Higgs boson 
self-coupling. In fact, a global analysis of the Higgs boson data collected so far reveals that non-standard decays 
of the Higgs boson (including invisible decays) with branching ratio (BR) upto $\sim 23\%$ are still consistent with 
experimental measurements \cite{Aad:2015pla}. 

The study of non-standard decay modes of the Higgs boson in various scenarios can thus be a good probe of 
new physics, lepton flavor violating (LFV) Higgs decays being one class of them.  
Among them decay rate of the channel, $h \rightarrow \mu \tau$ is relatively less constrained. The ATLAS collaboration 
has set an upper limit on BR($h \rightarrow \mu \tau$)$< 1.43 \%~$ at 95$\%$ confidence level with the run-I data 
collected at an integrated luminosity of 20.3 $\ifb$ \cite{Aad:2015gha}. At the same centre-of-mass energy, CMS has reported an upper 
limit of BR($h \rightarrow \mu \tau$)$< 1.51 \%$ at 95$\%$ confidence level with an integrated luminosity 
19.7 fb$^{-1}$ \cite{Khachatryan:2015kon}. 
The CMS collaboration have further updated their analysis with the $\sqrt{s} = 13$ TeV (run-II) data at an integrated luminosity 
2.3 $\ifb$ and puts an upper limit  BR($h \rightarrow \mu \tau$)$< 1.2 \%~$ \cite{CMS:2016qvi}. Side by side with these direct 
searches, several low-energy flavor violating processes, e.g. $\tau\to \mu\gamma$,  $\tau \rightarrow 3 \mu$, muon electric dipole moment (EDM), muon $(g-2)$ etc. 
put indirect constraints on the the Higgs flavor violating couplings \cite{Blankenburg:2012ex,Harnik:2012pb,Belusca-Maito:2016axk,Banerjee:2016foh}. 
In the context of specific models, attempts have been made to study this non-standard flavor violating decay for 
supersymmetric \cite{DiazCruz:1999xe,deLima:2015pqa,Arhrib:2012mg,Arhrib:2012ax,Abada:2014kba,Arganda:2015naa,
Arganda:2015uca,Aloni:2015wvn,Alvarado:2016par,Han:2000jz} as well as 
non-supersymmetric extensions of SM, including two Higgs doublet models \cite{DiazCruz:1999xe,
Crivellin:2015mga,Omura:2015nja,Dorsner:2015mja,Crivellin:2015hha,Botella:2015hoa,Arhrib:2015maa,Benbrik:2015evd}, the simplest 
little Higgs model \cite{Lami:2016mjf}, Randall-Sundrum scenarios \cite{Blanke:2008zb,Casagrande:2008hr}, and models containing
leptoquarks \cite{Cheung:2015yga} etc.

While further accumulation of data at the LHC 13 TeV run will be helpful in probing smaller BR($h \rightarrow \mu \tau$), 
the upper limit is not expected to improve in a drastic manner \cite{Banerjee:2016foh}. In this context, the relatively cleaner environment of 
electron-positron colliders can be more useful.  
We, therefore, explore the possibility of probing the same decay mode of the Higgs boson in an $e^+ e^-$ collider 
with the aim of improving upon the existing upper limit on its branching ratio imposed by the LHC. 

We have adopted a model-independent approach. In practice, such lepton flavor violating Higgs decays can happen in extensions of the single-doublet scenario, such as those considered in references \cite{Branco:2011iw,Das:2015kea}. In addition, terms originating from higher-dimensional operators which encapsulate physics at a high scale may drive such decays \cite{Grzadkowski:2010es, Buchmuller:1985jz, Harnik:2012pb}.

 It is obvious that the event rates for the ($\mu\tau$) final state 
depend, in addition to BR($h\to\mu\tau$), on the Higgs 
production rate in $e^+e^-$ collisions, where the $hVV$ ($V = W, Z$) interaction vertex is involved. We allow the 
possibility of new physics in $hVV$ coupling as well, as perhaps can be expected in a scenario that drives flavor violating Higgs decays in the leptonic sector. We do this by {\bf (i)} scaling the $hVV$ coupling strength, 
keeping the Lorentz structure same as SM, {\bf(ii)} introducing CP-even dimension-6 operators with new Lorentz structures. 
In the second scenario, momentum-dependent interactions can alter the kinematics 
of Higgs production. The existing constraints on such anomalous coupling have been taken into account \cite{Corbett:2012dm,Corbett:2012ja,Banerjee:2013apa, Banerjee:2012xc, Banerjee:2015bla}. 

The paper is oraganised as follows. In section \ref{sec:frame} we present the theoretical framework including two types of modifications at the production level as mentioned earlier. In this section 
we also discuss the relevant constraints derived from precision observables and their impact on the parameters characterizing physics beyond the Standard Model (BSM). Section \ref{Higgs_production} includes modification of Higgs production rates considering two aforementioned scenarios.
In section \ref{sec:coll} 
detailed collider simulation at different center-of-mass energies has been reported. We summarize and conclude in section \ref{sec:concl}.
%%%%%%%%%%%%%%%%%%%%%%%%%%%%%%%%%%%%%%%%%%%%%%%%%%%%%%%%%%%%%%%%%%%%%%%%%%%%%
\section{Scheme of the analysis}
\label{sec:frame}
%%%%%%%%%%%%%%%%%%%%%%%%%%%%%%%%%%%%%%%%%%%%%%%%%%%%%%%%%
The objective of this study is to examine the reach of $e^+ e^-$ colliders in probing the lowest possible BR$ (h \rightarrow \mu \tau)$, using a model-independent approach. For this, we study the different dominant Higgs production modes at different centre-of-mass energies and further decay of the Higgs boson to $\mu\tau$. Since the signal event rate depends on both Higgs production cross-section as well as its decay branching ratio, we explore the possibility of BSM physics in both production and decay. For the decay of Higgs in $\mu\tau$ mode, instead of introducing a specific kind of coupling, we adopt a model-independent approach where the corresponding branching ratio itself is varied upto the allowed limit. We further take into account both the leptonic and hadronic decays of $\tau$, resulting in various final states in order to do a comparative study. The final state in the leptonic $\tau-$decay consists of two opposite-sign same- or different-flavored leptons ($\mu \mu$ or $e \mu$) and $\me$. The hadronic decay ultimately leads to a $\mu + \tau_{had}(j) + \me$ final state. The Higgs mass is reconstructed from various observed decay products using the collinear approximation \cite{Ellis:1987xu}, which has been discussed later in section \ref{sec:coll}.

The dominant production channels of the Higgs boson at $e^+ e^-$ collision is
$e^+ ~ e^- \rightarrow Z ~ h$ at low center-of-mass energies such as $\sqrt{s} = 250~\rm{GeV}$. $e^+ ~ e^- \rightarrow h ~ \nu_e ~ \bar{\nu}_e$ driven by $W$-fusion dominates at $\sqrt{s} = 500~ \rm{GeV ~ and} ~ 1000 ~ \rm{GeV}$ (the production cross-section in $ZZ$ fusion is negligible). Therefore $hVV$ interaction
($V = W, Z$) is involved at the production level both at high and low energies.

We include new physics effects at the production level, by modifying the Standard Model $hVV$ couplings in two possible ways : 
\begin{itemize}
\item One can bring in just a multiplicative factor in the $hVV$ interactions.
\item The effect of various dimension-6 operators with new Lorentz structures in $hVV$ interactions may have some role to play.
\end{itemize}

Any change in the predicted values of Higgs couplings is bound to affect the electroweak precision data \cite{Corbett:2012ja,Banerjee:2013apa,Banerjee:2015bla} and the Higgs signal strengths in 
various decay modes. The allowed departure of the oblique electroweak parameters from their SM predicted values can be obtained from \cite{Baak:2014ora}: 
\bea
\Delta S = 0.05\pm 0.11 \,,
\Delta T = 0.09 \pm 0.13 \,,
\Delta U = 0.01 \pm 0.11 \,.
\label{STU}
\eea
The signal strength in a particular decay channel of Higgs boson is defined as,
%%%%%%%%%%%%%%%%%%%%%
\bea
\mu_{h \rightarrow X} &=& \frac{\sigma^{\rm{BSM}}(gg \rightarrow h) \times {\rm{BR}}^{\rm{BSM}}(h \rightarrow X)}
{\sigma^{\rm{SM}}(gg \rightarrow h)\times {\rm{BR}}^{\rm{SM}}(h \rightarrow X)} \,, \nonumber\\
 &=& \frac{\sigma^{\rm{BSM}}(gg \rightarrow h) \times \Gamma^{\rm{BSM}}(h \rightarrow X) \times \Gamma_{\rm{tot}}^{\rm{SM}}}
{\sigma^{\rm{SM}}(gg \rightarrow h)\times  \Gamma^{\rm{SM}}(h \rightarrow X) \times \Gamma_{\rm{tot}}^{\rm{BSM}}} \,.
\eea
%%%%%%%%%%%%%%%%%%%%%
$\sigma^{\rm{SM}}(gg \rightarrow h)$, ${\rm{BR}}^{\rm{SM}}(h \rightarrow X)$ being the production 
cross-section of Higgs boson via gluon-gluon fusion and the branching ratio of that particular decay mode $h \rightarrow X$ in the SM. 
$\sigma^{\rm{BSM}}(gg \rightarrow h)$, ${\rm{BR}}^{\rm{BSM}}(h \rightarrow X)$ are their BSM counterparts respectively. 

For the Higgs signal strength ($\mu$), we have used the combined results obtained from ATLAS and CMS 
\cite{TheATLASandCMSCollaborations:2015bln,CMS:2015kwa} derived from both $\sqrt{s} = 7$ TeV and $\sqrt{s} = 8 $ TeV run
of the LHC as shown in Table \ref{table2}. 
$2 \sigma $ allowed ranges for all the $\mu$-values have been used throughout our analysis.
%%%%%%%%%%%%%%%%%%%%%%%%%%%%%%%%%
\begin{table}[h]
\begin{center}
\begin{tabular}{|| c | c | c | c ||}
\hline
Decay mode & ATLAS  & CMS & ATLAS + CMS  \\ \hline
$\mu_{h \rightarrow \gamma \gamma}$ & $1.15^{+0.27}_{-0.25}$ & $1.12^{+0.25}_{-0.23}$ &  $1.16^{+0.20}_{-0.18}$ \\
$\mu_{h \rightarrow Z Z^*}$ & $1.51^{+0.39}_{-0.34}$ & $1.05^{+0.32}_{-0.27}$ & $1.31^{+0.27}_{-0.24}$ \\
$\mu_{h \rightarrow W W^*}$ & $1.23^{+0.23}_{-0.21}$ & $0.91^{+0.24}_{-0.21}$ & $1.11^{+0.18}_{-0.17}$ \\
$\mu_{h \rightarrow \tau \bar\tau}$ & $1.41^{+0.40}_{-0.35}$ & $0.89^{+0.31}_{-0.28}$ & $1.12^{+0.25}_{-0.23}$\\
$\mu_{h \rightarrow b \bar{b}}$ &$0.62^{+0.37}_{-0.36}$ & $0.81^{+0.45}_{-0.42}$ & $0.69^{+0.29}_{-0.27}$ \\
\hline
\end{tabular}
\caption{Signal strengths of different decay channels of Higgs boson obtained at $\sqrt{s} = 7$ TeV and $\sqrt{s} = 8 $ TeV.}
\label{table2}
\end{center}
\end{table}
%%%%%%%%%%%%%%%%%%%%%%%%%%%%%%%%%%%%%%%%%%%%%%%%%%%%%%%%%%%%%%%%%%%%%%%%%%%%%%%%
\section{Modification of Higgs production rates}
\label{Higgs_production}
\subsection{Modification of SM $hVV$ coupling with multiplicative factors only }
\label{sec:frame_hvv_mult}
%%%%%%%%%%%%%%%%%%%%%%%%%%%%%%%%%%%%%%%%%%%%%%%%%%%%%%%%%
Taking the Lorentz structure of the $hVV$ interaction to be same as the SM, the modified Lagrangian can be written as
\bea
\mathcal{L}_{eff}^{hVV} \supset a_W \left(\frac{2 m_W^2}{v}\right) h W_\mu^+ W^{\mu-} + a_Z \left(\frac{ m_Z^2}{v}\right) h Z_\mu Z^{\mu} \,
\label{SM_multiplicative}
\eea
where $a_W$, $a_Z$ are the multiplicative factors, $m_W$ and $m_Z$ are the masses of $W$ and $Z$ boson respectively and $v = 246$ GeV. It is assumed that Higgs couplings with the gluons and 
fermions are not modified with respect to the SM.

At $\sqrt{s} = 250~\rm{GeV}$, the dominant production process of the Higgs boson is $e^+ e^- \rightarrow Z h$ , 
which includes the $h Z  Z$ vertex, prompting us to vary $a_Z$. In a similar 
way, while considering $W$-fusion to be the dominant one among the production channels at 
$\sqrt{s} = 500~ \rm{GeV ~ and} ~ 1000 ~ \rm{GeV}$, multiplicative factor $a_W$ has been allowed to be varied, since the $W$-mediated channel $e^+ e^- \rightarrow \nu \bar{\nu} h$ dominates over the other production modes. 
Such scaling of the SM $hVV$ couplings arises, for example, when the SM Higgs doublet mixes with additional scalar multiplets. 
Any inequality of $a_W$ and $a_Z$ violates the invariance of custodial $SU(2)$ symmetry, resulting in tight constraints coming from the T-parameter \cite{Banerjee:2012xc,Corbett:2012ja}. The values of $a_W$ and 
$a_Z$ are also chosen consistently with the Higgs signal strengths.

While checking consistency with the LHC data it has been assumed that the Higgs boson is produced via gluon fusion which is the most efficient Higgs production mode at the LHC. Hence modification of the $hVV$ vertices does not affect the Higgs production cross-section. Thus the 
modifications in the $\mu$-values can be computed simply by the variation of Higgs branching ratios in 
different channels due to the introduction of the 
multiplicative factors $a_Z$ and $a_W$. \footnote{Note that, throughout this paper, while computing the modified 
$\mu$-values, we have considered Higgs boson production only via gluon fusion.} The variation of the known signal strengths due to non-vanishing BR$(h \rightarrow \mu \tau)$ is neglected. The obtained ranges of $a_Z$ and $a_W$ compatible with the above precision constraints are :
\bea
0.991 \leq  a_Z \leq 1.001 \,, 
0.997 \leq  a_W \leq 1.028 \,.
\eea

%%%%%%%%%%%%%%%%%%%%%%%%%%%%%%%%%%%%%%%%%%%%%%%%%%%%%%%%%%%%%%%%%%%%%%%%%%%%%%
\subsection{Modification of SM $hVV$ coupling by introducing dimension-6 operators}
\label{sec:frame_hvv_hdo}
%%%%%%%%%%%%%%%%%%%%%%%%%%%%%%%%%%%%%%%%%%%%%%%%%%%%%%%%%
We consider next the effect of introducing new Lorentz structures at the $hVV$ interaction vertices, 
keeping aforementioned multiplicative factors $a_Z$ and $a_W$ unity. For this purpose we have introduced the CP-even 
$SU(2)_L \times U(1)_Y$ invariant dimension-6 operators $\mathcal{O}_W$,$\mathcal{O}_{WW}$, 
$\mathcal{O}_B$ and $\mathcal{O}_{BB}$, as defined below \cite{Corbett:2012ja}: 
\bea
\mathcal{O}_{W} &=& \left(D_\mu \Phi \right)^\dag \hat{W}^{\mu \nu}\left(D_\nu \Phi \right) \,, \nonumber\\
\mathcal{O}_{WW} &=& \Phi^\dag \hat{W}_{\mu \nu} \hat{W}^{\mu \nu} \Phi \,, \nonumber\\
\mathcal{O}_{B} &=& \left(D_\mu \Phi \right)^\dag \hat{B}^{\mu \nu}\left(D_\nu \Phi \right) \,, \nonumber\\
\mathcal{O}_{BB} &=& \Phi^\dag \hat{B}_{\mu \nu} \hat{B}^{\mu \nu} \Phi \,.
\label{effective_operator}
\eea
with,
\bea
D_\mu \Phi &=& (\partial_\mu + \frac{i}{2} g_1 B_\mu + i g_2 \frac{\sigma_a}{2} W_\mu^a) \Phi \,, \nonumber\\
\hat{B}_{\mu \nu} &=& i \frac{g_1}{2} (\partial_\mu B_\nu - \partial_\mu B_\mu) \,, \nonumber\\
\hat{W}_{\mu \nu} &=& i \frac{g_2}{2} \sigma^a (\partial_\mu W_\nu^a - \partial_\nu W_\mu^a - g_2 f^{abc} W_\mu^b W_\nu^c) \,. \nonumber\\
\eea
Here $\Phi$ is SM- or SM-like scalar doublet, $g_1$ and $g_2$ are respectively the $U(1)_Y$ and $SU(2)_L$ gauge couplings, $\sigma_a$'s are the Pauli spin matrices and 
$f^{abc}$ are the $SU(2)$ structure constants. The operator $\mathcal{O}_{BW} = \Phi^\dag \hat{B}_{\mu \nu} \hat{W}^{\mu \nu} \Phi$ has 
been excluded, since it allows $Z$-$\gamma$ mixing at tree level, thereby violating the custodial $SU(2)$ symmetry which is 
responsible for keeping the $\rho$-parameter within its experimental bound \cite{Banerjee:2015bla, Corbett:2012ja}.
Hence the Lagrangian involving only $hVV$ interactions takes the form \cite{Corbett:2012ja}
\bea
\mathcal{L}_{eff}^{hVV} \supset \frac{f_W}{\Lambda^2} \mathcal{O}_W + \frac{f_{WW}}{\Lambda^2} \mathcal{O}_{WW} + \frac{f_B}{\Lambda^2} \mathcal{O}_B + 
\frac{f_{BB}}{\Lambda^2} \mathcal{O}_{BB} \,.
\label{L_eff}
\eea
where the $f_n$'s and $\Lambda$ are couplings and new physics scale respectively. We have taken $\Lambda = 1$ TeV 
throughout our analysis.

Since the $hVV$ couplings are modified in presence of these effective operators, it poses an apparent threat to perturbative unitarity in $V_L V_L \rightarrow V_L V_L (V = W, Z)$ at high energies. It should however be remembered that such a threat arises at scales above $\Lambda$, when additional degrees of freedom become operative. Unitarity is then expectedly ensured by the scenario which is responsible for such degrees of freedom.

The Lagrangian involving new Lorentz structures in $hVV$ interactions can be written as \cite{Corbett:2012ja},
\bea
\mathcal{L}_{eff}^{hVV} &=& g_{h \gamma \gamma} h A_{\mu \nu} A^{\mu \nu} + g_{h Z \gamma}^{(1)} A_{\mu \nu} Z^\mu \partial^\nu h 
+ g_{hZ\gamma}^{(2)} h A_{\mu \nu} Z^{\mu \nu} + g_{hZZ}^{(1)} Z_{\mu \nu} Z^\mu \partial^\nu h  \nonumber\\
&& + g_{h Z Z}^{(2)} h Z_{\mu \nu} Z^{\mu \nu} + g_{hWW}^{(1)} (W_{\mu \nu}^+ W^{-\mu} \partial^\nu h + {\rm{h.c.}}) + g_{hWW}^{(2)} h W_{\mu \nu}^+ W^{- \mu \nu} \,.
\eea
with effective couplings $g_{h\gamma\gamma}$, $g_{hZ\gamma}^{(1)}$, $g_{hZ\gamma}^{(2)} $, $g_{h Z Z}^{(1)}$, $g_{h Z Z}^{(2)}$, $g_{h W W}^{(1)}$ ,
$g_{h W W}^{(2)}$. Here $V_{\mu \nu} = \partial_\mu V_\nu - \partial_\nu V_\mu$ with $V = A, Z, W$. These effective couplings can be expressed as linear combination of the $f_n$'s, mentioned earlier in eq.(\ref{L_eff}).
\bea
g_{h\gamma\gamma} &=& -\left(\frac{g_2^2 v s_W^2}{2 \Lambda^2}\right)\frac{f_{BB} + f_{WW}}{2} \,, \nonumber\\
g_{hZ\gamma}^{(1)} &=&  \left(\frac{g_2^2 v}{2 \Lambda^2}\right) \frac{s_W (f_W - f_B)}{2 c_W} \,, \nonumber\\
g_{hZ\gamma}^{(2)} &=& \left(\frac{g_2^2 v}{2 \Lambda^2}\right) \frac{s_W ( s_W^2 f_{BB} -  c_W^2 f_{WW})}{ c_W} \,, \nonumber\\
g_{h Z Z}^{(1)} &=& \left(\frac{g_2^2 v}{2 \Lambda^2}\right) \frac{c_W^2 f_W + s_W^2 f_B}{2 c_W^2} \,, \nonumber\\
g_{h Z Z}^{(2)} &=& -\left(\frac{g_2^2 v}{2 \Lambda^2}\right) \frac{s_W^4 f_{BB} + c_W^4 f_{WW} }{2 c_W^2} \,, \nonumber\\
g_{h W W}^{(1)} &=& \left(\frac{g_2^2 v}{2 \Lambda^2}\right) \frac{f_W}{2} \,, \nonumber\\
g_{h W W}^{(2)} &=& -\left(\frac{g_2^2 v}{2 \Lambda^2}\right) f_{WW} \,.
\label{effective_couplings}
\eea
$c_W$ and $s_W$ are the short-hand notations for cos$\theta_W$ and sin$\theta_W$ respectively, $\theta_W$ being the Weinberg angle. Here Higgs-gluon-gluon and Higgs-fermion-fermion interactions are taken to be same as the SM.

%The $hVV$ interaction terms involving derivatives of the gauge fields and %Higgs field give rise to momentum dependent amplitudes and 
%therefore can modify the kinematics in Higgs boson production via $e^+ e^- %\rightarrow Z h$ as well as $WW$-fusion. The decay of the Higgs 
%boson into weak bosons also gets modified and accordingly affects the signal strength calculation. In fact, LHC data \cite{TheATLASandCMSCollaborations:2015bln,CMS:2015kwa} puts the most stringent constraints on the $f_n$'s through 
%various signal strength measurements, as will be discussed in the next subsection.
%%%%%%%%%%%%%%%%%%%%%%%%%%%%%%%%%%%
%\subsubsection{Calculation of signal strengths in presence of higher %dimensional operators}
%\label{sec:hdm_ops}
%%%%%%%%%%%%%%%%%%%%%%%%%%%%%%%%%%%
For simplicity, we have switched on only one of the aforementioned four operators at a time. It is clear from Table \ref{table1} that $hZZ$ couplings are modified for non-zero $f_B$, $f_{W}$ and $f_{BB}$, $f_{WW}$ respectively. Likewise 
$g_{h W W}^{(1)}$ and $g_{h W W}^{(2)}$ depend on $f_W$ and $f_{WW}$ respectively. 
%%%%%%%%%%%%%%%%%%%%%%%%%%%%%%%%%%%%%%%%%%%%
\begin{table}[h!]
\begin{center}
\begin{tabular}{ | c | c|}
\hline
 Non-zero $f_n$'s & Modified couplings in eq.(\ref{effective_couplings}) \\ \hline 
 $f_B$ & $g_{hZ\gamma}^{(1)}$ , $g_{h Z Z}^{(1)}$ \\  
 $f_{BB}$  &  $g_{h\gamma\gamma}$ , $g_{hZ\gamma}^{(2)}$, $g_{h Z Z}^{(2)}$ \\ 
 $f_W$ & $g_{hZ\gamma}^{(1)}$, $g_{h Z Z}^{(1)}$, $g_{h W W}^{(1)}$ \\
 $f_{WW}$ & $g_{h\gamma\gamma}$ , $g_{hZ\gamma}^{(2)}$, $g_{h Z Z}^{(2)}$, $g_{h W W}^{(2)}$ \\
 \hline 
\end{tabular}
\caption{Modified couplings for non-zero $f_n$'s (taken one at a time)}
\label{table1}
\end{center}
\end{table}
%%%%%%%%%%%%%%%%%%%%%%%%%%%%%%%%%%%%%%%%%%%%

Thus the partial decay widths for the channels $h \rightarrow Z Z^*$, $h \rightarrow W W^*$, $h \rightarrow \gamma\gamma$  
and $h \rightarrow Z \gamma$ 
are expected to be modified for non-zero $f_n$'s. The modified partial decay width of the Higgs boson can be expressed as 
polynomials of the effective coupling constants, {\em{i.e.}} $f_B$ , $f_{BB}$, $f_W$, $f_{WW}$, partial width of 
all the other channels being same as the SM. Since the decay width of $h \rightarrow Z \gamma$ is rather small in SM, its modification 
will hardly change the final results. Thus we have not included modification of this particular decay width, nor do we include the decay width for $h \rightarrow \mu \tau$ which contributes not more than $1\%$ 
to the total Higgs decay rate. Expressions for modified decay widths involving the four effective couplings are as follows :
\begin{itemize}
\item {\underline{Involving $f_B$ only}} : \\
%%%%%%%%%%%%%%%%%%%%%%%%%%%%%%%%%%%
\bea
\Gamma_{h \rightarrow Z Z^*}^{\rm{BSM}} &=& 1.0745 \times 10^{-4} - 3.205 \times 10^{-7} f_B + 1.751 \times 10^{-9} f_{B}^2 \,, 
%\Gamma_{h \rightarrow Z \gamma}^{\rm{BSM}} &=& 8.896 \times 10^{-5} - 3.205 %\times 10^{-7} f_B + 1.751 \times 10^{-9} f_{B}^2 \,. (CHECK)
\label{mod_sigstr_fB}
\eea
%%%%%%%%%%%%%%%%%%%%%%%%%%%%%%%%%%%
%%%%%%%%%%%%%%%%%%%%%%%%%%%%%%%%%%%
\item {\underline{Involving $f_{BB}$ only}} : \\
%%%%%%%%%%%%%%%%%%%%%%%%%%%%%%%%%%%
\bea
\Gamma_{h \rightarrow Z Z^*}^{\rm{BSM}} &=& 1.0745 \times 10^{-4} + 3.458 \times 10^{-8} f_{BB} + 2.435 \times 10^{-10} f_{BB}^2 \,, \nonumber \\
\Gamma_{h \rightarrow \gamma \gamma}^{\rm{BSM}} &=& 9.279 \times 10^{-6} + 1.675 \times 10^{-5} f_{BB} + 6.691 \times 10^{-6} f_{BB}^2 \,,
%\Gamma_{h \rightarrow Z \gamma}^{\rm{BSM}} &=& 1.048 \times 10^{-5} + 1.675 %\times 10^{-5} f_{BB} + 6.691 \times 10^{-6} f_{BB}^2 \,. (CHECK)
\label{mod_sigstr_fBB}
\eea
%%%%%%%%%%%%%%%%%%%%%%%%%%%%%%%%%%%
%%%%%%%%%%%%%%%%%%%%%%%%%%%%%%%%%%%
\item {\underline{Involving $f_W$ only}} : \\
%%%%%%%%%%%%%%%%%%%%%%%%%%%%%%%%%%%
\bea
\Gamma_{h \rightarrow Z Z^*}^{\rm{BSM}} &=& 1.0745 \times 10^{-4} - 1.0103 \times 10^{-6} f_W + 1.075 \times 10^{-8} f_W^2 \,, \\
\Gamma_{h \rightarrow W W^*}^{\rm{BSM}} &=& 8.7505\times 10^{-4} - 9.99 \times 10^{-6} f_W + 1.8604 \times 10^{-8} f_W^2 \,, 
%\Gamma_{h \rightarrow Z \gamma}^{\rm{BSM}} &=& 1.048 \times 10^{-5} + 1.675 %\times 10^{-5} f_{BB} + 6.691 \times 10^{-6} f_{BB}^2 \,. (CHECK)
\eea
%%%%%%%%%%%%%%%%%%%%%%%%%%%%%%%%%%%
%%%%%%%%%%%%%%%%%%%%%%%%%%%%%%%%%%%
\item {\underline{Involving $f_{WW}$ only}} : \\
%%%%%%%%%%%%%%%%%%%%%%%%%%%%%%%%%%%
\bea
\Gamma_{h \rightarrow Z Z^*}^{\rm{BSM}} &=& 1.0745 \times 10^{-4} + 4.452 \times 10^{-7} f_{WW} + 6.838 \times 10^{-10} f_{WW}^2 \,, \\
\Gamma_{h \rightarrow \gamma \gamma}^{\rm{BSM}} &=& 9.279 \times 10^{-6} + 7.66 \times 10^{-6} f_{WW} + 5.599 \times 10^{-6} f_{WW}^2 \,, \\
\Gamma_{h \rightarrow W W^*}^{\rm{BSM}} &=& 8.7505 \times 10^{-4} + 8.484 \times 10^{-6} f_{WW} + 2.2 \times 10^{-8} f_{WW}^2 \,, 
%\Gamma_{h \rightarrow Z \gamma}^{\rm{BSM}} &=& 6.3665 \times 10^{-6} + %1.0035 \times 10^{-5} f_{WW} + 3.82 \times 10^{-6} f_{WW}^2 \,. (CHECK)
\eea
%%%%%%%%%%%%%%%%%%%%%%%%%%%%%%%%%%%
%%%%%%%%%%%%%%%%%%%%%%%%%%%%%%%%%%%
\end{itemize}

The $f_n$-independent term as well as those linear and quadratic in $f_n$ in the above equations correspond to contributions from SM, 
interference between SM and BSM, and purely BSM respectively.
For each case the modifications in the $\mu$-values have been calculated to compare with the existing constraints.

The allowed ranges of $f_B$, $f_W$, $f_{BB}$, $f_{WW}$ have been derived using $2\sigma$-allowed ranges of the electroweak precision 
observables as given in eq.(\ref{STU}) and $2\sigma$-allowed ranges of the signal strength values shown in Table~\ref{table2}. The allowed 
ranges for the individual couplings are given in Table~\ref{table0}. In presence of $f_{BB}$ and $f_{WW}$, $\Gamma_{h \rightarrow \gamma \gamma}$ gets modified. The partial decay width $\Gamma_{h \rightarrow \gamma \gamma}$ becomes minimum at $f_{BB} = -1.25$ and $f_{WW} = -0.68$ respectively (taking one of them non-zero at a time). In the intermediated excluded region around the minimum, signal strength of the channel $h \rightarrow \gamma \gamma$ becomes lower than its $2\sigma$ allowed lower limit. Thus the intermediate region $-2.38 < f_{BB} < -0.12$ for $f_{BB}$ and $-1.04 < f_{WW} < -0.319$ for $f_{WW}$ are excluded by the $2\sigma$ constraint on the signal strength.
%%%%%%%%%%%%%%%%%%%%%%%%%%%%%%%%%%%%
%\begin{table}[h!]
%\begin{center}
%\begin{tabular}{ | c | c|}
%\hline
%Couplings & Allowed ranges \\ \hline 
% $f_B$ & [-3.4 : 11.0] \\  
% $f_{BB}$  &  [-2.78 : -2.5] $\cup$ [-0.126 : 0.283] \\ 
% $f_W$ &  [-5.8 : 14.5]\\
% $f_{WW}$ & [ -1.86 : -1.04] $\cup$ [ -0.319 : 0.5]\\
% \hline 
%\end{tabular}
%\caption{Allowed ranges of $f_n$'s with $\Lambda = 1$ TeV obtained by using $1\sigma$-allowed ranges of the electroweak precision 
%observables and $2\sigma$-allowed ranges of the Higgs signal strengths.}
%\label{table0}
%\end{center}
%\end{table}
\begin{table}[h!]
\begin{center}
\begin{tabular}{ | c | c|}
\hline
Couplings & Allowed ranges \\ \hline 
 $f_B$ & [-11.74 , 18.66] \\  
 $f_{BB}$  &  [-2.78 , -2.38] $\cup$ [-0.12 , 0.283] \\ 
 $f_W$ &  [-25.1 , 25.8]\\
 $f_{WW}$ & [ -1.86 , -1.04] $\cup$ [ -0.319 , 0.5]\\
 \hline 
\end{tabular}
\caption{Allowed ranges of $f_n$'s with $\Lambda = 1$ TeV obtained by using $2\sigma$-allowed ranges of the electroweak precision 
observables and $2\sigma$-allowed ranges of the Higgs signal strengths.}
\label{table0}
\end{center}
\end{table}
%%%%%%%%%%%%%%%%%%%%%%%%%%%%%%%%%%%%%
%%%%%%%%%%%%%%%%%%%%%%%%%%%%%%%%%%%%%%%%%%%%%%%%%%%%%%%%%%%
\section{Collider Analysis}
\label{sec:coll}
%%%%%%%%%%%%%%%%%%%%%%%%%%%%%%%%%%%%%%%%%%%%%%%%%%%%%%%%%%%
The prospect of observing LFV decays of the 125 GeV Higgs boson has been explored in the context of the LHC 
\cite{Aad:2015gha,Aad:2016blu,Khachatryan:2015kon,Banerjee:2016foh, Bhattacherjee:2015sia}. 
These studies indicate that the smallest LFV decay branching ratio (BR($h\to\mu\tau$)) that can be probed at the 
high-luminosity run of the LHC at 14 TeV is $\sim 10^{-2}$. A recent phenomenological study \cite{Barradas-Guevara:2017ewn} provides the lower bound of the branching ratio of $h \rightarrow \mu \tau$ to be $\sim 10^{-3}$. A lepton collider on the other hand provides a much cleaner 
environment and thus provides ideal platform to probe such non-standard decays of the Higgs boson \cite{Banerjee:2016foh,Chakraborty:2016gff}. 
Our primary aim 
in this section would be to assess whether one can probe even smaller branching ratios with different center-of-mass energies. At $\sqrt{s}=250$ GeV, $e^+e^-\to Zh$ is the most dominant production mode 
of the Higgs boson. However, this production cross-section diminishes with increasing center-of-mass energy unlike the 
$W$-fusion channel, $e^+e^-\to h\nu_e\bar\nu_e$. As a result, at $\sqrt{s}=500$ and $1000$ GeV, the $W$-fusion channel 
turns out to be the dominant contributor in Higgs production (production cross-section of $ZZ$-fusion is negligible even at high $\sqrt{s}$). We have explored the search prospects of the present 
scenario at all these three center-of-mass energies. 

In order to perform our collider analysis, the new interaction vertices have been included in \texttt{FeynRules} \cite{Christensen:2008py,Alloul:2013bka}. We 
have used MadGraph5 \cite{Alwall:2011uj,Alwall:2014hca} to generate events at the parton level and subsequently \texttt{Pythia-6} \cite{Sjostrand:2006za} for decay, 
showering and hadronisation. While generating the events, we have used the default dynamic factorisation and renormalisation 
scales \cite{madgraph_scale} at MadGraph. Detector simulation has been performed using \texttt{Delphes-3.3.3} \cite{deFavereau:2013fsa,Selvaggi:2014mya,Mertens:2015kba}. 
Jets have been reconstructed with FastJet \cite{Cacciari:2011ma} using anti-kt \cite{Cacciari:2008gp} algorithm. We have taken the $\tau$-tagging efficiency and the probability of a jet faking $\tau$ to be 60$\%$
and 2$\%$ respectively.  
In order to identify the leptons, photons and jets in the final state, we have imposed the following primary selection criteria :
\begin{itemize}
\item All the charged leptons are selected with a minimum transverse momentum cut-off 10 GeV, {\em i.e.} $p_T^\ell > 10$ GeV. Further, the electrons and muons must also lie within the pseudo-rapidity window 
$|\eta^{e}| < 2.5$ and $|\eta^{\mu}| < 2.5$ respectively.
\item All the photons are selected with $p_T^\gamma > 10$ GeV and $|\eta^{\gamma}| < 2.5$.
\item All the jets in the final state must satisfy $p_T^j > 30$ GeV and $|\eta^{j}| < 2.5$.
\item It is ensured that the final state particles are well separated by demanding $\Delta R > 0.4$ between lepton-jet pairs and $\Delta R > 0.25$ between lepton pairs.
\end{itemize}
Let us first consider the scenario described in section~\ref{sec:frame} where we have used the maximally allowed 
value of $a_Z$ ($ a_Z = 1.001$) and $a_W$ ($ a_W = 1.028$) in agreement with the electroweak precision observables and 
Higgs signal strength measurements, in order to determine the cross-sections in $e^+e^-\to Z h$ and 
$e^+e^-\to \nu_e\bar\nu_e h$ production modes respectively. Later in this section, we proceed to discuss the possible improvement 
in the results in presence of higher-dimensional operators. 
%%%%%%%%%%%%%%%%%%%%%%%%%%%%%%%%
\subsection{$e^+e^-\to Z h$ at $\sqrt{s}=250$ GeV}
\label{sec:eezh250}
%%%%%%%%%%%%%%%%%%%%%%%%%%%%%%%%
For this production mode, we chose to study the cleaner channel where the $Z$-boson decays leptonically. Further, 
we have considered both the leptonic and hadronic decays of the $\tau$ arising from the 125 GeV $h$ decay. 
Thus depending on the decays of $\tau$, the various final states can be as follows.

\begin{itemize}
 \item Tau decaying leptonically : $4\ell + \me , \ell = e, \mu$
 \begin{enumerate}
  \item $e^+ e^- \rightarrow Z h , Z \rightarrow \mu^+ \mu^- , h \rightarrow \mu \tau \rightarrow e \mu + \me  $ \\
  $\Rightarrow$ $e + 3 \mu + \cancel{E}$
  \item $e^+ e^- \rightarrow Z h , Z \rightarrow \mu^+ \mu^- , h \rightarrow \mu \tau \rightarrow 2 \mu + \me  $ \\
  $\Rightarrow$  $ 4 \mu + \me$
  \item  $e^+ e^- \rightarrow Z h , Z \rightarrow e^+ e^- , h \rightarrow \mu \tau \rightarrow e \mu + \me  $ \\
  $\Rightarrow$ $\mu + 3 e + \me$
  \item $e^+ e^- \rightarrow Z h , Z \rightarrow e^+ e^- , h \rightarrow \mu \tau \rightarrow 2 \mu + \me  $ \\
  $\Rightarrow$ $ 2 e + 2 \mu + \me$
 \end{enumerate}
 \item Tau decaying hadronically : $3\ell + \tau_{had} + \me , \ell = e, \mu$
 \begin{enumerate}
 \item $e^+ e^- \rightarrow Z h , Z \rightarrow \mu^+ \mu^- , h \rightarrow \mu \tau \rightarrow  \mu \tau_{had} + \me  $ \\
  $\Rightarrow$ $ \tau_{had} + 3 \mu + \me$
   \item $e^+ e^- \rightarrow Z h , Z \rightarrow e^+ e^- , h \rightarrow \mu \tau \rightarrow \mu \tau_{had} + \me $ \\
  $\Rightarrow$ $ 2 e +  \mu + \tau_{had} + \me$
  \end{enumerate}
\end{itemize}

The corresponding major SM backgrounds can arise from the following channels :
\begin{enumerate}
 \item $e^+ e^- \rightarrow Z h $ , $Z \rightarrow \ell \bar{\ell}$ , $h \rightarrow \ell \bar{\ell}$ ; $\ell = e, \mu, \tau$
 \item $e^+ e^- \rightarrow Z Z $ , $Z \rightarrow \ell \bar{\ell}$ , $Z \rightarrow \ell \bar{\ell}$ ; $\ell = e, \mu, \tau$
 \item $e^+ e^- \rightarrow Z Z \ell \ell$ , $Z \rightarrow \ell \bar{\ell}$ , $Z \rightarrow \nu \bar{\nu}$; $l = e, \mu, \tau$ ; $ \nu = \nu_e , \nu_\mu , \nu_\tau$
 \item $e^+ e^- \rightarrow Z Z \nu \bar{\nu}$ , $Z \rightarrow \ell \bar{\ell}$ , $Z \rightarrow \ell \bar{\ell}$ ; $ \nu = \nu_e , \nu_\mu , \nu_\tau$
\end{enumerate}

%%%%%%%%%%%%%%%%%%%%%%%%%%
\begin{itemize}
\item {\bf Final state: $4\ell + \me$ :} \\
%%%%%%%%%%%%%%%%%%%%%%%%%%
We have used the following set of cuts to identify our signal events and reduce the SM background contribution to get the best 
possible signal to background ratio.
\begin{itemize}
\item {\bf A0 :} The final state must consist of four leptons with at least one $\mu$. A veto has been applied on the 
jets in the final state since the $\tau$ in this case is expected to decay leptonically.
\item{\bf A1 :} For such a signal, some amount of missing energy is always expected to arise from Higgs boson decay. A normalized 
distribution of $\me$ for the signal process as well as the most dominant background channels $ZZ$ and $Zh$ are shown in Fig.\ref{fig:me} 
where the blue line corresponds to the signal process and the black and the red lines correspond to $ZZ$ and $Zh$ background production 
channels respectively.   
%%%%%%%%%%%%%%%%%%%%%%%%%%
\begin{figure}[h!]
\begin{center}
\includegraphics[scale=0.45]{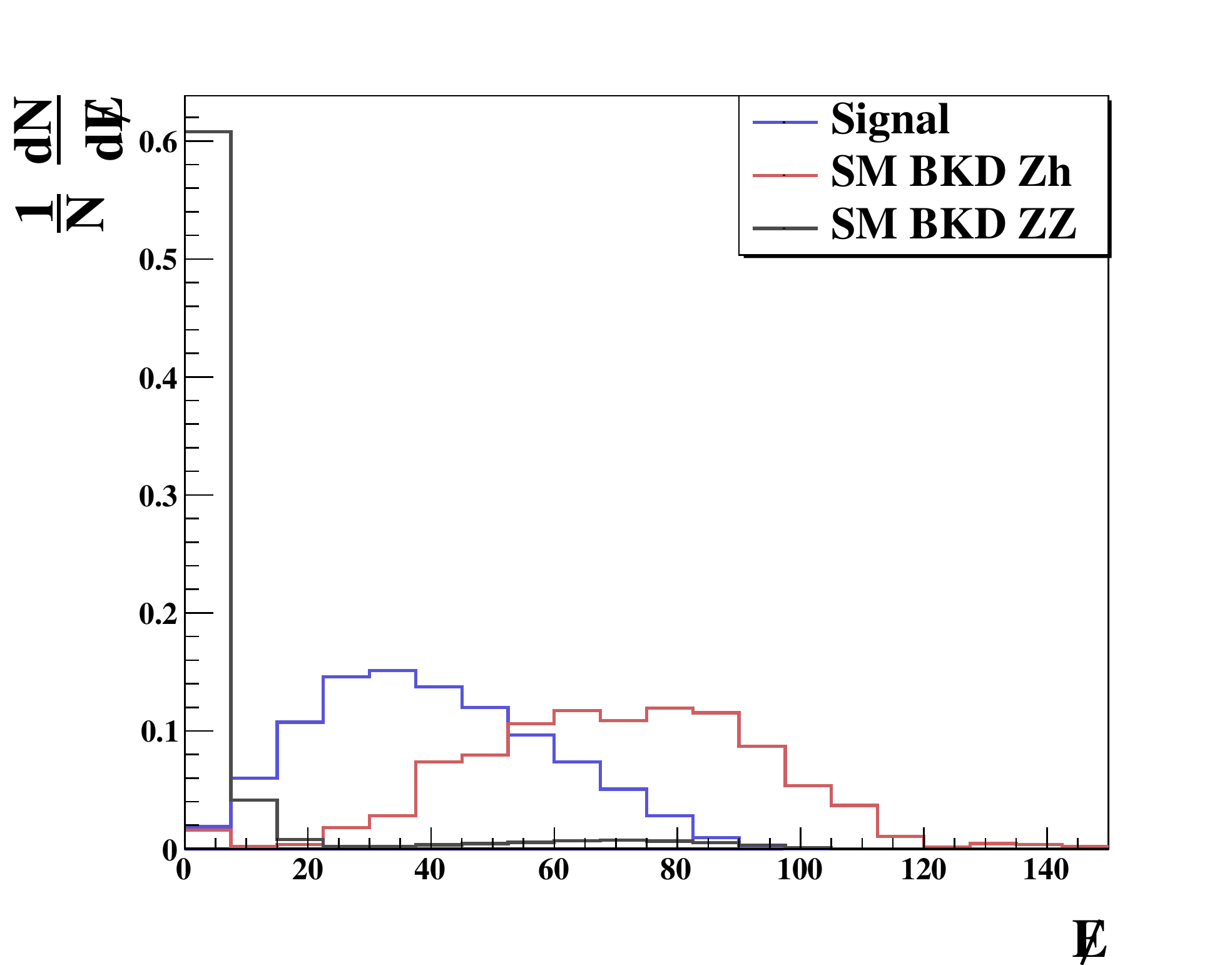}
\caption{Normalized $\me$ distribution for signal and backgrounds at $\sqrt{s}=250$ GeV for final state: $4\ell + \me$.}
\label{fig:me}
\end{center}
\end{figure}
%%%%%%%%%%%%%%%%%%%%%%%%%%
We demand, $100~{\rm GeV} > \me > 20~{\rm GeV}$.
\item {\bf A2 :} At least one of the same-flavor opposite-sign lepton pairs are expected to arise from the $Z$-boson decay 
in the signal. Hence all such pairs have been identified in order to reconstruct their invariant masses ($M_{\ell\ell}$) 
and the pair for which $M_{\ell\ell}$ lies closest to the $Z$-boson mass ($m_Z$) has been identified. We have 
then demanded that $|M_{\ell\ell} - m_Z| < 10$ GeV for that particular pair of same-flavor opposite-sign leptons.
\item {\bf A3 :} Once the leptons arising from the $Z$-boson decays are identified, the rest of the leptons and missing energy 
should mostly originate from the decay of $h$. In order to reconstruct the Higgs mass, the collinear approximation \cite{Ellis:1987xu} has been used as mentioned earlier. The mass of Higgs being much greater than that of $\tau$, the decay products of $\tau$ are highly boosted in its original direction. Thus the direction of the neutrino momenta can be approximated to be in the same direction of the visible decay products of $\tau$. Thus the transverse component of the neutrino momentum can be estimated by taking the projection of the of the missing transverse energy in the direction of the visible tau decay products, {\em i.e.} 
$\vec{p}_T^{\nu} = \vec{\met}.\hat{p}_T^{\tau_{\rm vis}}$.

We have used the collinear mass 
($M_{\rm coll}$) \cite{CMS:2014hha}, defined as 
\begin{eqnarray}
M_{\rm coll} = \frac{M_{\rm vis}}{\sqrt{x_{\tau_{vis}}}} \,.
\end{eqnarray}   
where $M_{\rm vis}$ represents the invariant mass of the remaining leptons and the fraction of the tau momentum carried by the visible tau decay products is 
$x_{\tau_{\rm vis}}=\frac{|\vec{p}_T^{\tau_{\rm vis}}|}{|\vec{p}_T^{\tau_{\rm vis}}| + |\vec{p}_T^{\nu}|}$. 
 
 Fig.\ref{fig:mcoll} represents the normalized distribution of $M_{\rm coll}$ 
for the signal and background channels with the same color coding as in Fig.\ref{fig:me}. The signal clearly shows a sharper 
peak around the Higgs boson mass region. 
\end{itemize}
%%%%%%%%%%%%%%%%%%%%%%%
\begin{figure}[h!]
\begin{center}
\includegraphics[scale=0.45]{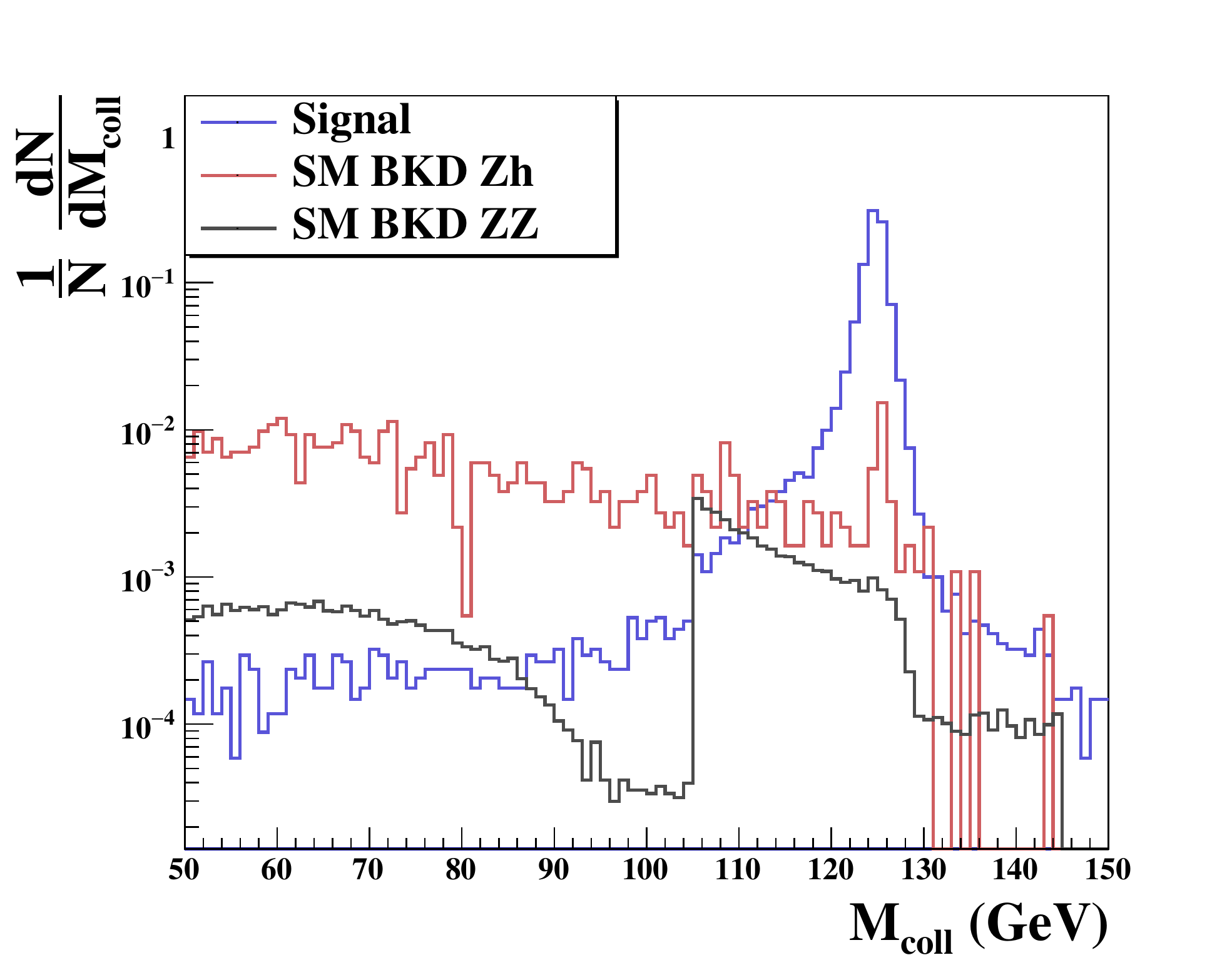}
\caption{Normalized $M_{\rm coll}$ distribution for signal and backgrounds at $\sqrt{s}=250$ GeV for final state: $4\ell + \me$.}
\label{fig:mcoll}
\end{center}
\end{figure}
%%%%%%%%%%%%%%%%%%%%%%
We have demanded that, $(m_h + 20)~{\rm GeV} > M_{\rm coll} > (m_h - 20)~{\rm GeV}$. 

In Table~\ref{tab:zh_4l_250} we have presented the detailed cut-flow numbers obtained from our collider simulation at 
$\sqrt{s}=250$ GeV for integrated luminosity ${\mathcal{L}}=250~\ifb$ corresponding to the signal $4\ell + \me$ (with 
$BR(h \rightarrow \mu \tau) = 9.78 \times 10^{-3}$) as well as the different SM background channels. 
%%%%%%%%%%%%%%%%%%%%%%%%%%%%%%%%%%%%%%%%%%%%%%%%%%%%n{table}[ht!]
%%\scriptsize
\begin{table}[h]
\begin{center}
\begin{tabular}{||c|c|c|c|c|c||}
\hline
\multicolumn{1}{||c|}{\bf Process} &
\multicolumn{5}{|c||}{\bf $\sqrt{s}=$250 GeV} \\ 
\cline{2-6}
\multicolumn{1}{||c|}{} &
\multicolumn{1}{|c|}{$\sigma$ (pb)} & 
\multicolumn{4}{|c||}{NEV ($\mathcal{L}=$250 $\ifb$)} \\ 
\cline{3-6}
 &  & {\bf A0} & {\bf A1} & {\bf A2} & {\bf A3}  \\
\hline
$e^+ e^- \rightarrow Z h$  & $1.37 \times 10^{-4}$ &  5 & 4  & 4 & 4  \\
\hline\hline
$e^+ e^- \rightarrow Z h $ & 0.24 & 27 & 23  & 20 & 2 \\
\hline
$e^+ e^- \rightarrow Z Z $  & $9.48 \times 10^{-3}$ &  515 & 25 & 20 & - \\
\hline
$e^+ e^- \rightarrow Z Z l l$  & $2.558 \times 10^{-4}$ & 1  & - & - & -\\
\hline
$e^+ e^- \rightarrow Z Z \nu {\bar{\nu}} $  & $1.3 \times 10^{-3}$ &  1 & 1 & - &  - \\
\hline
\end{tabular}
\caption{Cross-sections of the signal and various background channels for leptonic decay of $\tau$, are shown in pb alongside the number of expected events for the 
individual channels at 250 $\ifb$ luminosity after applying each of the 
cuts {\bf A0} - {\bf A3} as listed in the text. NEV $\equiv$ number of events. Signal cross-section has been quoted for BR$(h \rightarrow \mu \tau) = 9.78 \times 10^{-3}$.}
\label{tab:zh_4l_250}
\end{center}
\end{table}%
%%%%%%%%%%%%%%%%%%%%%%%%%%%%%%%%%%%%%%%%%%%%%%%%%%%%%

As evident from Table~\ref{tab:zh_4l_250}, the $ZZ$ production channel is potentially the most dominant contributor to the SM background. However, 
the $\me$ ({\bf A1}) and $M_{\rm coll}$ ({\bf A3}) cuts turn out to be particularly effective in reducing this background. The SM $Zh$ production 
channel also can be a possible source of background due to its large production cross-section, but the signal requirement of multiple leptons and no associated jets reduces this contribution which is further dented by the cut {\bf A3}. Clearly, the signal rate being extremely small, one requires 
a large integrated luminosity in order to observe any such events. As the numbers in Table~\ref{tab:zh_4l_250} indicates, one would need an 
integrated luminosity of $\approx 450~\ifb$ in order to gain a $3\sigma$ statistical significance for this signal at $\sqrt{s}=250$ GeV with our choice 
of BR$(h \rightarrow \mu \tau) = 9.78 \times 10^{-3}$.   
%%%%%%%%%%%%%%%%%%%%%%%%%%
\item {\bf Final state: $3\ell + 1\tau{\rm -jet} + \me$ :} \\
%%%%%%%%%%%%%%%%%%%%%%%%%%
As discussed earlier, such final states may arise if the $\tau$ originating from the Higgs decays hadronically. 
We have used the following set of cuts to identify our signal events and reduce the SM contribution to get the best
possible signal to background ratio.
\begin{itemize}
\item {\bf B0 :} The final state must consist of three leptons with at least one $\mu$. We further demand that the number of jets in the final state should be restricted to one and it must be identified as a $\tau$-jet. 
\item{\bf B1 :} For a hadronic decay of the $\tau$, the $\me$ distribution is softer compared to the leptonic 
decay scenario. This is indicated by Fig.~\ref{fig:me_had} which shows the normalized distribution of $\me$ for the $3\ell + 1\tau-{\rm jet} + \me$ 
final state for the signal as well as $ZZ$ and $Zh$ background production channels with the same color coding as Fig.~\ref{fig:me}.
%%%%%%%%%%%%%%%%%%%%%%%%%
\begin{figure}[h!]
\begin{center}
\includegraphics[scale=0.45]{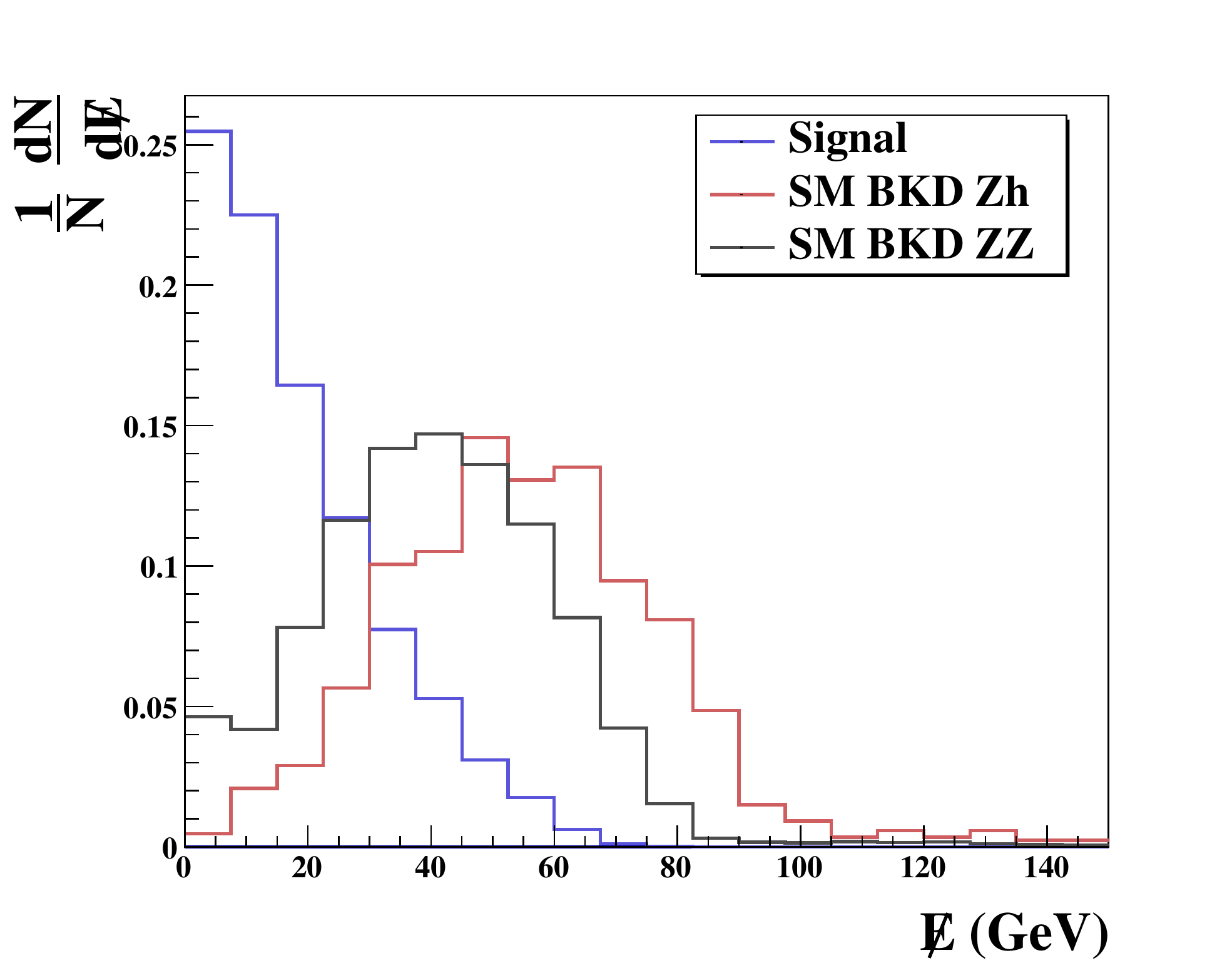}
\caption{Normalized $\me$ distribution for signal and backgrounds at $\sqrt{s}=250$ GeV for final state: $3\ell + 1\tau-{\rm jet} + \me$.}
\label{fig:me_had}
\end{center}
\end{figure}
%%%%%%%%%%%%%%%%%%%%%%%%

We, therefore, demand a missing energy upper limit: $ \me < 30~{\rm GeV} $.  
\item {\bf B2 :} If the other two leptons in the event apart from the one $\mu$ originating from $h$ happen to 
be electrons, they have most likely been originated from the $Z$-boson.  
However, if all the three leptons in the event happen to be muons, we follow the same exercise as described in 
{\bf A2} to identify the $\mu^+ \mu^-$ pair originating from the $Z$-boson and similarly restrict the resulting $M_{\ell\ell}$ within
$|M_{\ell\ell} - m_Z| < $10 GeV.
\item {\bf B3 :} In this case, the visible decay products of the Higgs boson consist of a lepton and a $\tau$-jet. 
We reconstruct $M_{\rm coll}$ in a similar way as described in {\bf A3} and subsequently demand that, 
$(m_h + 20)~{\rm GeV} > M_{\rm coll} > (m_h - 20)~{\rm GeV}$. Fig.~\ref{fig:mcoll_had} represents the distribution of 
$M_{\rm coll}$ before applying the cuts.
%%%%%%%%%%%%%%%%%%%%%%%%%%%%%
\begin{figure}[h!]
\begin{center}
\includegraphics[scale=0.45]{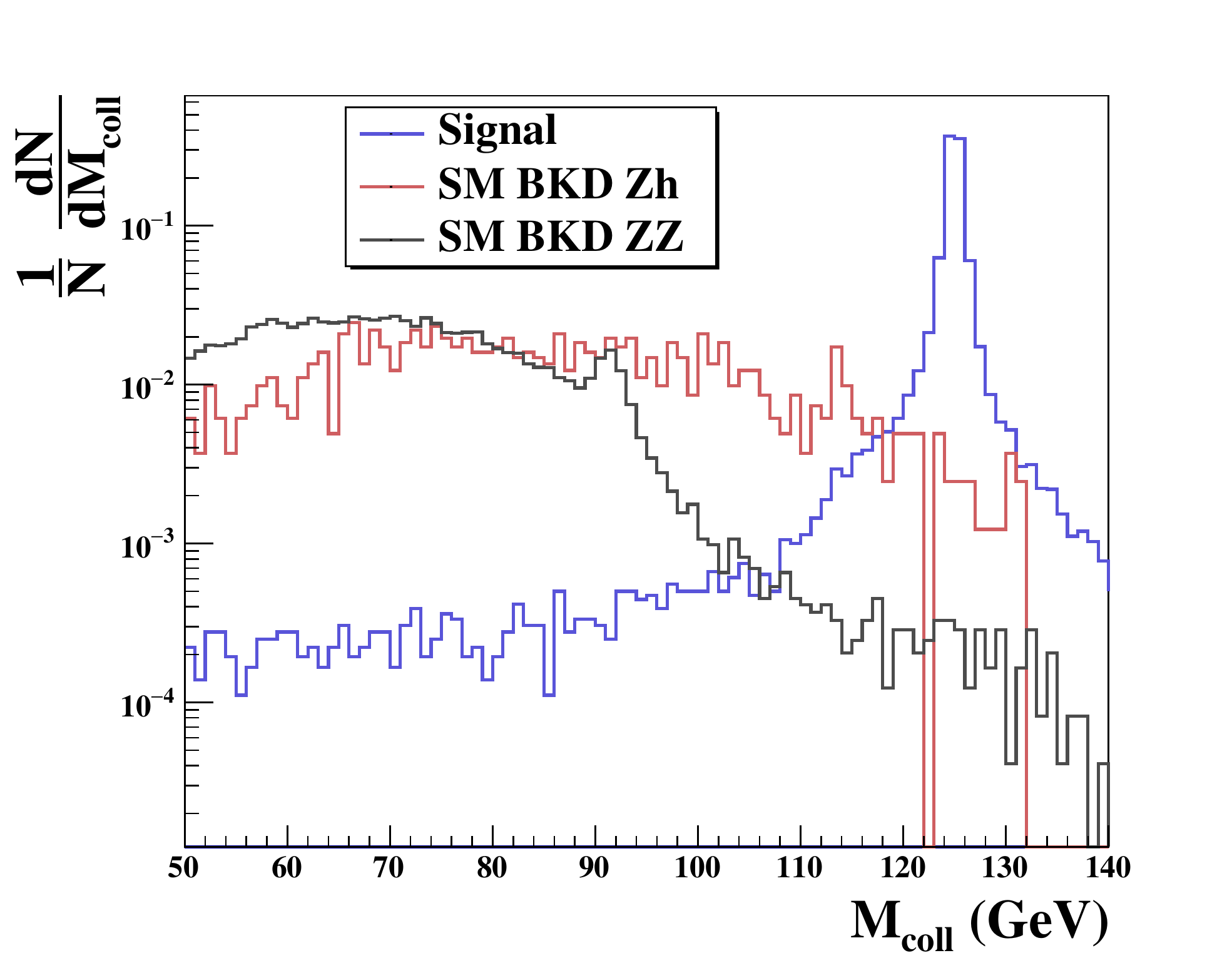}
\caption{Normalized $M_{\rm coll}$ distribution for signal and backgrounds at $\sqrt{s}=250$ GeV for final state: $3\ell + 1\tau-{\rm jet}+ \me$.}
\label{fig:mcoll_had}
\end{center}
\end{figure}
%%%%%%%%%%%%%%%%%%%%%%%%%%%%%%
\end{itemize}
In Table~\ref{tab:zh_3l1j_250} below we have presented the cut-flow numbers obtained from our collider simulation at 
$\sqrt{s}=250$ GeV and an integrated luminosity of ${\mathcal{L}}=250~\ifb$ corresponding to our signal $3\ell + 1\tau-{\rm jet}+ \me$ (with 
$BR(h \rightarrow \mu \tau) = 9.78 \times 10^{-3}$) as well as the different SM background channels. 
%%%%%%%%%%%%%%%%%%%%%%%%%%%%%%%%%%%%%%%%%%%%%%%%%%%%n{table}[ht!]
%%\scriptsize
\begin{table}[h]
\begin{center}
\begin{tabular}{||c|c|c|c|c|c||}
\hline
\multicolumn{1}{||c|}{\bf Process} &
\multicolumn{5}{|c||}{\bf $\sqrt{s}=$250 GeV} \\ 
\cline{2-6}
\multicolumn{1}{||c|}{} &
\multicolumn{1}{|c|}{$\sigma$ (pb)} & 
\multicolumn{4}{|c||}{NEV ($\mathcal{L}=$250 $\ifb$)} \\ 
\cline{3-6}
 &  & {\bf B0} & {\bf B1} & {\bf B2} & {\bf B3}  \\
\hline
$e^+ e^- \rightarrow Z h$ & $1.37 \times 10^{-4}$ & 5 &  3 & 3 & 3 \\
\hline\hline
 $e^+ e^- \rightarrow Z h$ & 0.24 & 10 & 1  & 1 & 1 \\
\hline
$e^+ e^- \rightarrow Z Z$ & $9.48 \times 10^{-3}$ & 25 & 6 & 6 & - \\
\hline
$e^+ e^- \rightarrow Z Z l l$ & $2.558 \times 10^{-4}$ & - & - & - & - \\
\hline
$e^+ e^- \rightarrow Z Z \nu {\bar{\nu}}$ & $1.3 \times 10^{-3}$ & -  & - & - & - \\
\hline
\end{tabular}
\caption{Cross-sections of the signal and various background channels for hadronic decay of $\tau$,
are shown in pb alongside the number of expected events for the 
individual channels at 250 $\ifb$ luminosity after applying each of the 
cuts {\bf B0} - {\bf B3} as listed in the text. NEV $\equiv$ number of events. 
Signal cross-section has been quoted for BR$(h \rightarrow \mu \tau) = 9.78 \times 10^{-3}$.}
\label{tab:zh_3l1j_250}
\end{center}
\end{table}%
%%%%%%%%%%%%%%%%%%%%%%%%%%%%%%%%%%%%%%%%%%%%%%%%%%%%%

As evident from Table~\ref{tab:zh_3l1j_250}, the $ZZ$ production channel is potentially the dominant contributor to the SM background, However, 
in this case also, $\me$ ({\bf B1}) and $M_{\rm coll}$ ({\bf B3}) cuts turn out to be particularly effective in reducing this background. The SM 
$Zh$ production channel also can be possible source of background which is reduced effectively by {\bf B1}. As the numbers indicate, much like 
the leptonic $\tau$-decay scenario, here also one requires an integrated luminosity of $\approx 450~\ifb$ in order to obtain a $3\sigma$ statistical 
significance with a choice of BR$(h \rightarrow \mu \tau) = 9.78 \times 10^{-3}$.  

In Table~\ref{lowest_br_az} we have shown the lowest possible reach of $e^+ e^-$ collider in probing BR$(h\to\mu\tau)$ at the $3 \sigma$ level for different integrated luminosities for the 
two possible final states, $3\ell + 1\tau-{\rm jet}+ \me$ and $4\ell + \me$ studied at $\sqrt{s}=250$ GeV for comparison. 
%%%%%%%%%%%%%%%%%%%%%%%%%%%%%%%%%%%%%%%%
\begin{table}
\begin{center}
\begin{tabular}{|| c | c | c | c ||}
\hline
$\mathcal{L} (fb^{-1})$ & lowest BR in $(4\ell+\me)$  & lowest BR in $(3\ell+\tau_{had}+\me)$ & Combined BR  \\ \hline
350 & 0.0109 & 0.0111 &  $7.42 \times 10^{-3}$ \\
500 & $8.87 \times 10^{-3}$ & $8.96 \times 10^{-3}$ & $6.0 \times 10^{-3}$\\
1000 & $5.94 \times 10^{-3}$ & $5.92 \times 10^{-3}$  & $4.09 \times 10^{-3}$  \\
\hline
\end{tabular}
\caption{Lowest branching ratio that can be probed with 3$\sigma$ statistical significance for the two different final states (arising from leptonic and hadronic decay of $\tau$) at $\sqrt{s}$=250 GeV with BR$(h \rightarrow \mu \tau) = 9.78 \times 10^{-3}$. The last column indicates the BR reach when the event rates of these two final states are combined together.}
\label{lowest_br_az}
\end{center}
\end{table}
%%%%%%%%%%%%%%%%%%%%%%%%%%%%%%%%%%%%%%%%
We have presented the numbers for three predicted luminosities, {\em i.e.} $350~\ifb$, $500~\ifb$ and 
$1000~\ifb$ at 3$\sigma$ significance \footnote{The statistical significance ($\sigma'$) has been calculated for the number 
of signals ($s$) and number of backgrounds ($b$) using $ \sigma' = \sqrt{2 [(s+b) {\rm ln} (1+ \frac s b) - s]}$.}. Results for 
both hadronic and leptonic decay modes of $\tau$ have been quoted individually alongwith the combined result (obtained by 
merging the results from two different decay modes of $\tau$). Both the leptonic and hadronic decay modes of $\tau$ perform with similar effectiveness in probing the lowest 
possible BR($h \rightarrow \mu \tau$) at $\sqrt{s} = 250$ GeV. The result obtained by combining the two different final states, however, can do slightly better 
than the individual channels as indicated by the numbers in the last column of Table~\ref{lowest_br_az}. It can be inferred that the lowest probed branching ratio at $\sqrt{s}=250$ GeV is $\sim 10^{-3}$.
%%%%%%%%%%%%
\end{itemize}
%%%%%%%%%%%%%%%%%%%%%%%%%%%%%%%%
\subsection{$e^+e^-\to \nu_e\nu_e h$ at $\sqrt{s}=1000$ GeV}
\label{sec:eennh}
%%%%%%%%%%%%%%%%%%%%%%%%%%%%%%%%
The $W$-fusion production mode, namely $e^+ e^-\to\nu_e\bar\nu_eh$, although having a negligible cross-section compared to 
$e^+ e^-\to Zh$ at $\sqrt{s}=250$ GeV, becomes the most dominating one at $\sqrt{s}=500$ GeV and 1000 GeV. 
The production cross-section in the channel $e^+ e^-\to Zh$, on the other hand, starts gradually decreasing beyond 
$\sqrt{s}=250$ GeV and thus becomes less relevant for $\sqrt{s}=500$ GeV or above. It would be interesting to 
see if a further increase in the centre-of-mass energy can help us reach better sensitivity in probing a smaller $h\to\mu\tau$ 
branching ratio. The $W$-fusion production mode gives rise to a single Higgs associated with two electron neutrinos 
that contribute to the missing energy. Hence depending on the leptonic or hadronic decay of the $\tau$, the final state may consist of the following signal channels:
\begin{itemize}
 \item Tau decaying leptonically :$2\ell + \me , \ell = e, \mu$
 \begin{enumerate}
  \item $e^+ e^- \rightarrow \nu_e {\bar{\nu}_e} h ,  h \rightarrow \mu \tau \rightarrow e \mu + \me $ \\
  $\Rightarrow$ $e +  \mu + \me$
  \item $e^+ e^- \rightarrow \nu_e {\bar{\nu}_e} h , h \rightarrow \mu \tau \rightarrow 2 \mu + \me  $ \\
  $\Rightarrow$  $ 2 \mu + \me$
 \end{enumerate}
 \item Tau decaying hadronically :  $\mu + \tau_{had} + \me$
 \begin{enumerate}
 \item $e^+ e^- \rightarrow \nu_e {\bar{\nu}_e} h  , h \rightarrow \mu \tau \rightarrow  \mu \tau_{had} + \me  $ \\
  $\Rightarrow$ $ \tau_{had} +  \mu + \me$
  \end{enumerate}
\end{itemize}

The relevant SM background channels consist of $W^+W^-$, $\tau^+\tau^-$, $t\bar t$, $ZZ$, $Zh$, $W^+W^-Z$, $ZZZ$, 
$t\bar tZ$, $ZZh$ and $\ell jj\nu$. Our analysis with $\sqrt{s}=500$ GeV reveals that there is little scope to increase the sensitivity in probing 
$BR(h\to\mu\tau)$ to much smaller values than what we have already obtained for the $\sqrt{s}=250$ GeV case with 
$e^+ e^-\to Zh$ production mode even at higher ($\approx 1000~\ifb$) luminosities. At $\sqrt{s} = 500$ GeV, the overall rate of the Higgs production through $e^+ e^- \rightarrow \nu_e \bar{\nu_e} h$ and its subsequent decay to $\mu \tau$ is of the order of $10^{-4}$ pb. Moreover, the background coming from $W^+ W^-$ channel dominates over the other SM backgrounds at this center-of-mass energy. The number of background events coming from $W^+ W^-$ channel being very large compared to the number of signals even after applying suitable cuts on the kinematic variables, makes it non-trivial to achieve a $3\sigma$ significance. Hence we chose not to present the 
numerical results from this simulation. Instead we have presented below the results obtained for the 
$\sqrt{s}=1000$ GeV analysis, where the production rate is considerably higher.
%%%%%%%%%%%%%%%%%%%%%%%%%%%%%%%%%%%%%%%%%%%%%
\begin{itemize}
\item {\bf Final state: $2\ell + \me$ :} \\
%%%%%%%%%%%%%%%%%%%%%%%%%%%%%%%%%%%%%%%%%%%%%%%%%%%%%
Here we have used the following set of kinematical cuts in order to reduce the SM background contributions to gain 
best possible signal to background ratio.
\begin{itemize}
\item {\bf C0 :} There must be one hard muon along with another lepton (electron or muon) in the final state. 
Since the $\tau$ decays leptonically, there are no direct sources of jets. Hence we put a veto on jets on 
the final state including $\tau$- and $b$-jets.
\item {\bf C1 :} Missing energy distribution for the final state $2\ell + \me$ is shown in Fig.\ref{fig:me_lep_1000} for 
the signal events (blue line) as well as the dominant background production channels, namely, $t\bar t$ (brown line), 
$WW$ (black line), $WWZ$ (violet line) and $ZZ$ (grey line) at $\sqrt{s}=1000$ GeV.  
%%%%%%%%%%%%%%%%%%%%%%%%%%%%%%%%%%%%%%%%%%
\begin{figure}[h!]
\begin{center}
\includegraphics[scale=0.45]{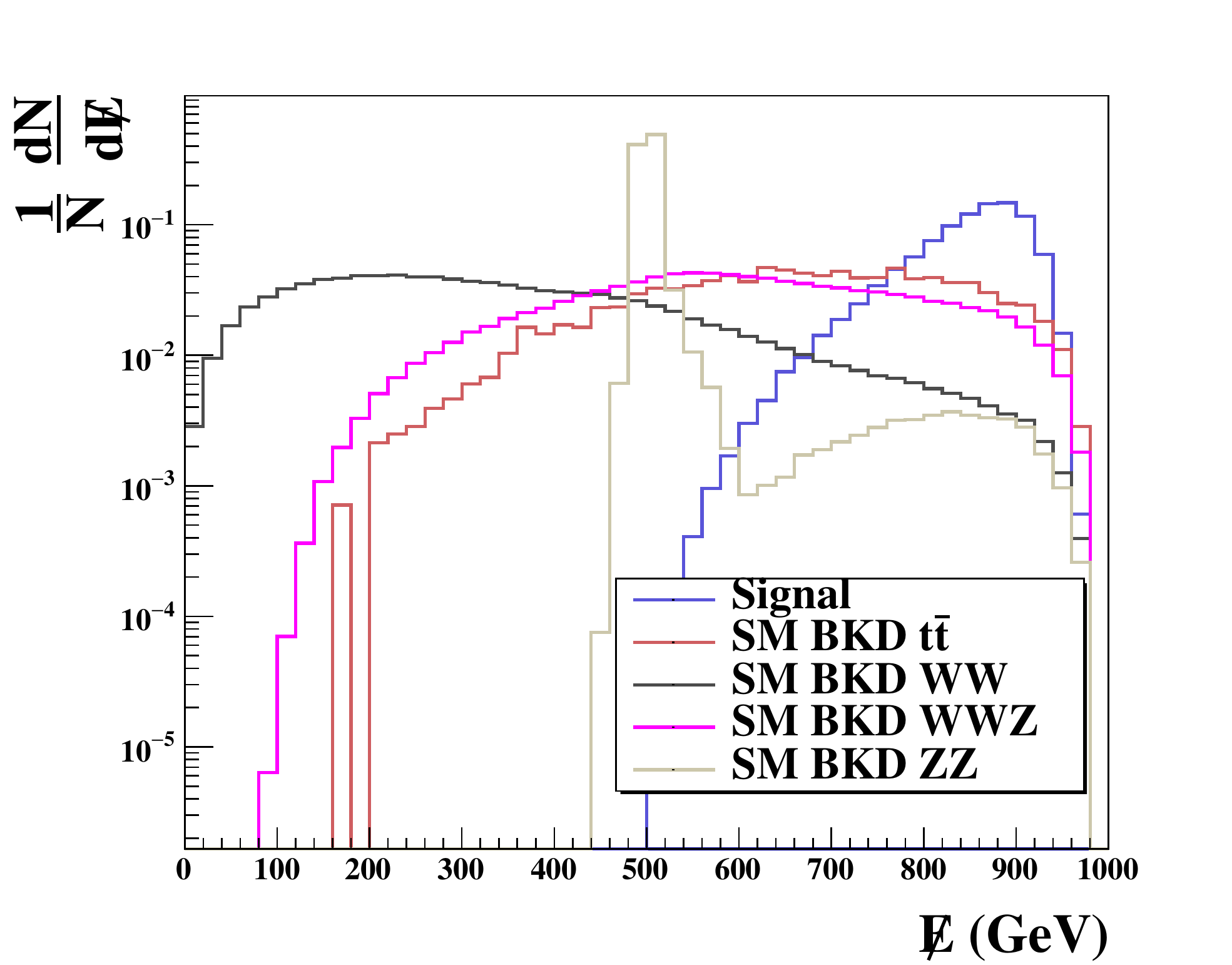}
\caption{Normalized $\me$ distribution for signal and backgrounds at $\sqrt{s}=1000$ GeV for final state : $2\ell + \me$.}
\label{fig:me_lep_1000}
\end{center}
\end{figure}
%%%%%%%%%%%%%%%%%%%%%%%%%%%%%%%%%%%%%%%%%

We demand a $\me$ window: $1000~{\rm GeV} > \me > 600~{\rm GeV}$.  
\item {\bf C2 :} In the signal events, both the leptons in the event are expected to arise from the Higgs decay 
whereas for the background events, two leptons can originate from two different parent particles and may have 
a larger angle in between them. For example, in the $W^+W^-$ background channel, the two leptons in the event 
are back to back and thus have a large separation angle which can be exploited to reduce the background 
contribution. This kinematic feature can be observed in Fig.~\ref{fig:costhe_lmu_1000} where the normalized distribution of 
${\rm cos}\theta_{\ell\mu}$ is shown for the signal and SM background events with the same color coding as in Fig.~\ref{fig:me_lep_1000}. 
%%%%%%%%%%%%%%%%%%%%%%%%%%%%%%%%%
\begin{figure}[h!]
\begin{center}
\includegraphics[scale=0.45]{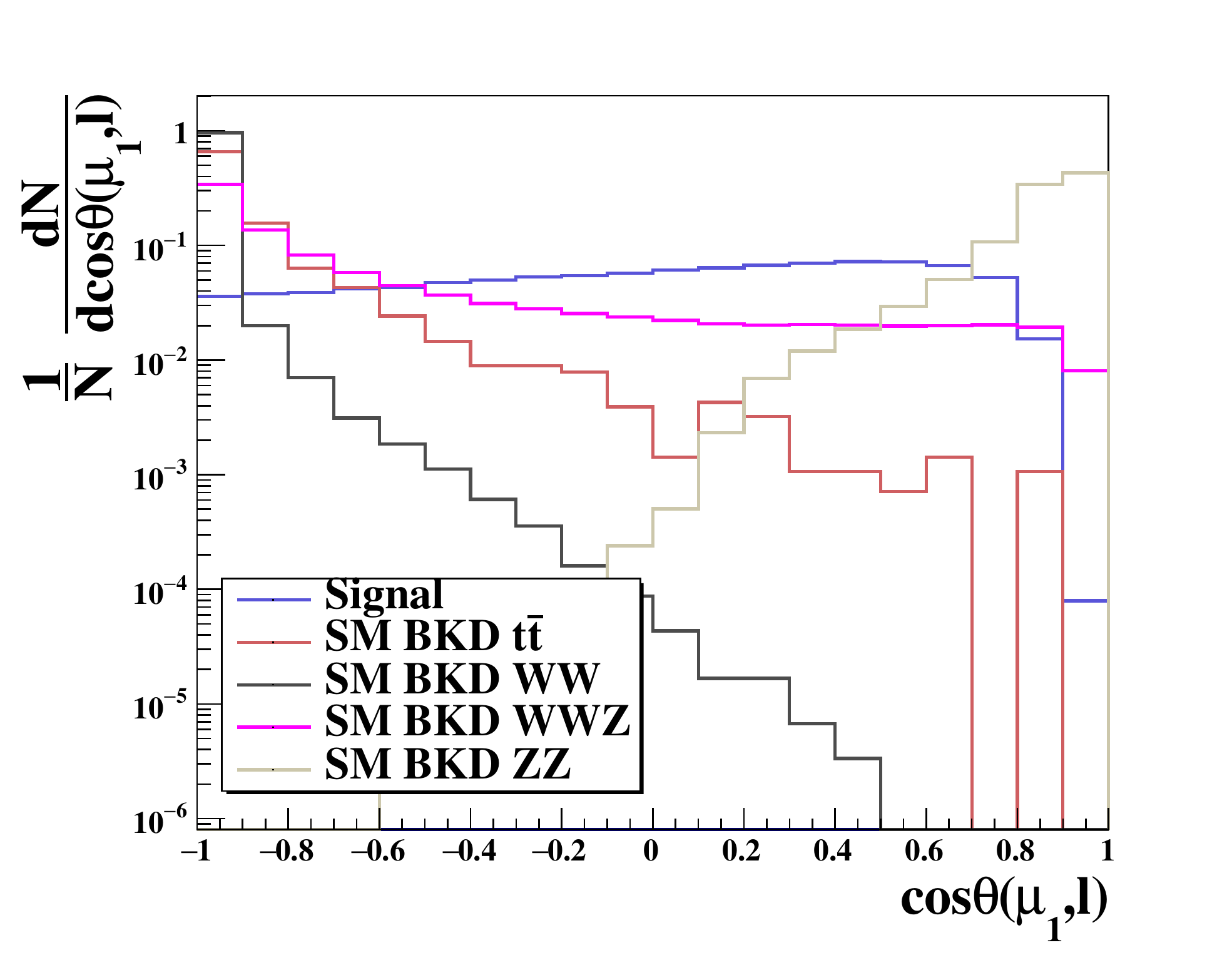}
\caption{Normalized ${\rm cos}\theta_{\ell\mu}$ distribution for signal and backgrounds at $\sqrt{s}=1000$ GeV for final state: $2\ell + \me$.}
\label{fig:costhe_lmu_1000}
\end{center}
\end{figure}
%%%%%%%%%%%%%%%%%%%%%%%%%%%%%%%%

We demand $0.9 > {\rm cos}\theta_{\ell\mu} > -0.8$.   
\item {\bf C3 :} We demand that the invariant mass of the visible particles, that is of the two-lepton system 
should lie within the region $120~{\rm GeV} > M_{\ell\mu} > 40~{\rm GeV}$. Fig.\ref{M_vis_1000_lep} represents normalized distribution 
of $M_{\ell\mu}$ for the signal and SM background events with the same color coding as in Fig.~\ref{fig:me_lep_1000}.
%%%%%%%%%%%%%%%%%%%%%%%%%%%%%%%%
\begin{figure}[h!]
\begin{center}
\includegraphics[scale=0.45]{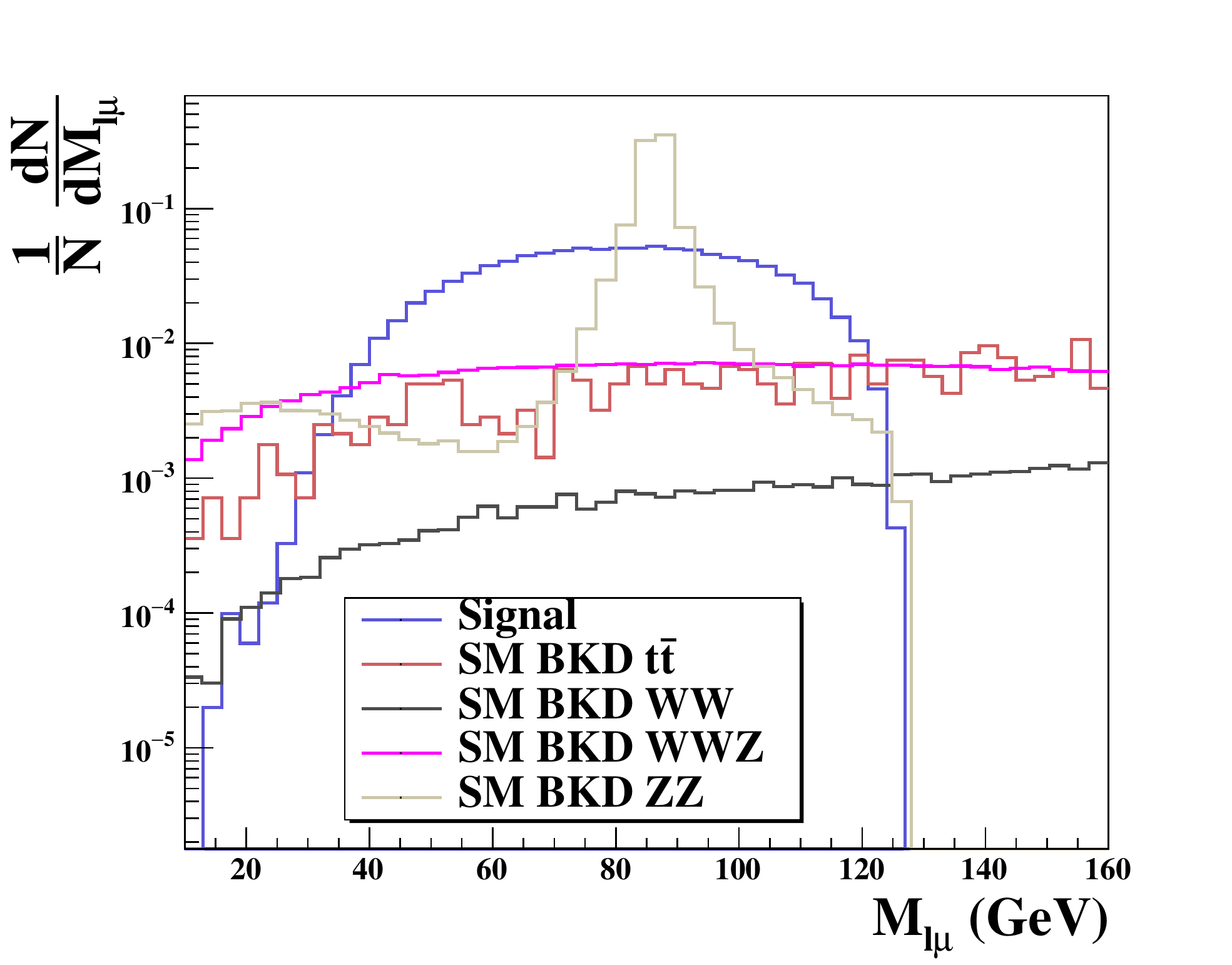}
\caption{Normalized $ M_{\ell\mu}$ distribution for signal and backgrounds at $\sqrt{s}=1000$ GeV for final state: $2\ell + \me$.}
\label{M_vis_1000_lep}
\end{center}
\end{figure}
%%%%%%%%%%%%%%%%%%%%%%%%%%%%%%%
\item {\bf C4 :} In our signal events, the hardest muon ($\mu_1$) is likely to be generated directly from the Higgs decay. 
Hence, we expect the missing energy vector, $\vec{\me}$ to be well separated from this muon. We demand, 
$3.14 > \Delta \phi(\mu_1,\vec{\me}) > 1.0$. Fig.\ref{del_phi_mume_lep} shows  the distribution of $\Delta \phi(\mu_1,\vec{\me})$
for the signal and SM background events with the same color coding as in Fig.~\ref{fig:me_lep_1000}.
%%%%%%%%%%%%%%%%%%%%%%%%%%%%%%
\begin{figure}[h!]
\begin{center}
\includegraphics[scale=0.45]{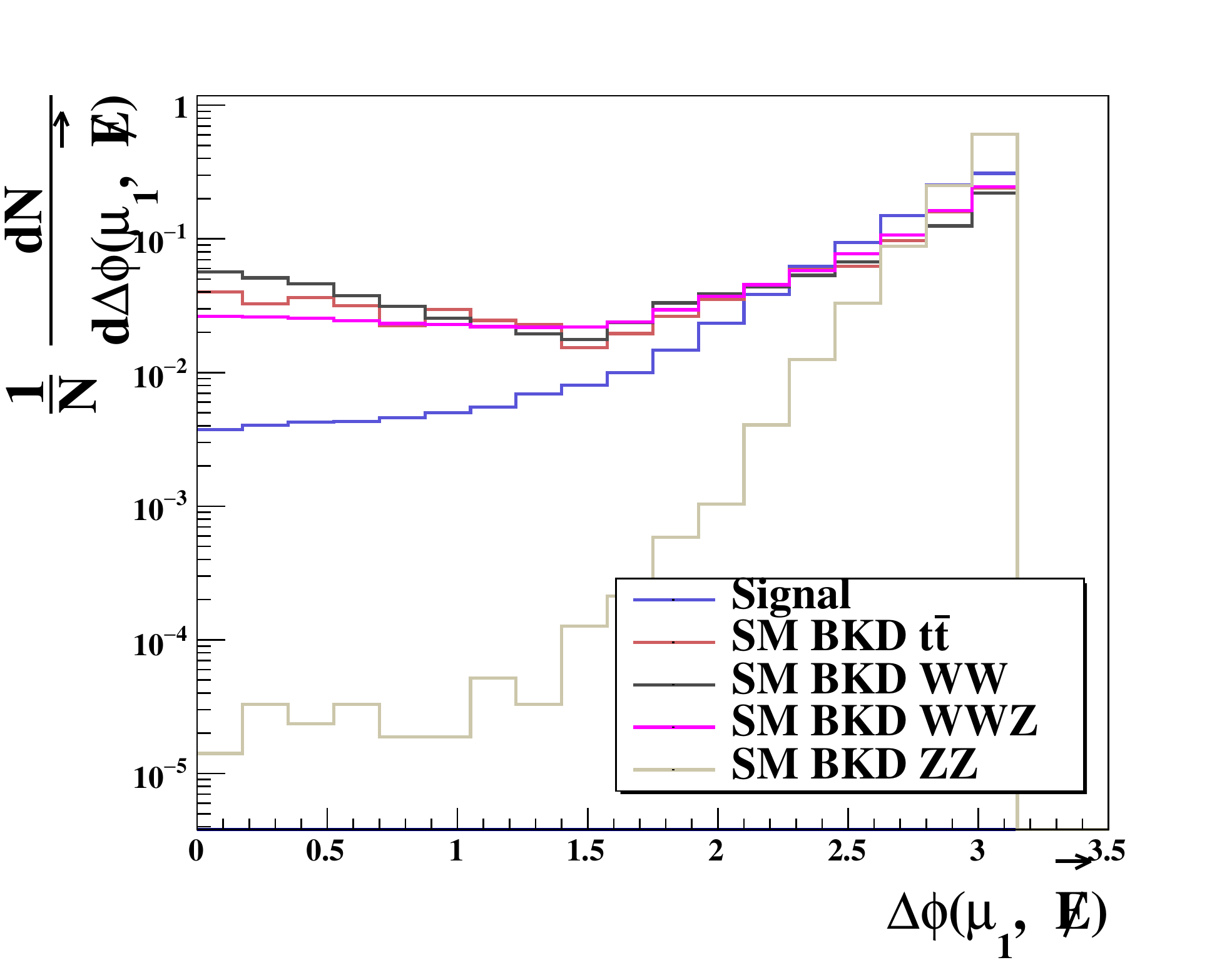}
\caption{Normalized $ \Delta \phi(\mu_1,\vec{\me})$ distribution for signal and backgrounds at $\sqrt{s}=1000$ GeV for final state: $2\ell + \me$.}
\label{del_phi_mume_lep}
\end{center}
\end{figure}
%%%%%%%%%%%%%%%%%%%%%%%%%%%%%%
\end{itemize}
In Table~\ref{tab:vvh_2l_1000}, we have presented the cut-flow numbers obtained from our simulation at 
$\sqrt{s}=1000$ GeV and an integrated luminosity of ${\mathcal{L}}=500~\ifb$ corresponding to our signal $2\ell + \me$ (with 
$BR(h \rightarrow \mu \tau) = 9.78 \times 10^{-3}$) as well as the different SM background channels.  
%%%%%%%%%%%%%%%%%%%%%%%%%%%%%%%%%%%%%%%%%%%%%%%%%%%%n{table}[ht!]
%%\scriptsize
\begin{table}[h]
\begin{center}
\begin{tabular}{||c|c|c|c|c|c|c||}
\hline
\multicolumn{1}{||c|}{\bf Process} &
\multicolumn{6}{|c||}{\bf $\sqrt{s}=$1000 GeV} \\
\cline{2-7}
\multicolumn{1}{||c|}{} &
\multicolumn{1}{|c|}{$\sigma$ (pb)} &
\multicolumn{5}{|c||}{NEV ($\mathcal{L}=$500 $\ifb$)} \\
\cline{3-7}
 &  & {\bf C0} & {\bf C1} & {\bf C2} & {\bf C3} & {\bf C4}  \\
\hline
$e^+ e^- \rightarrow \nu_e\bar{\nu}_e h$  &$2.01 \times 10^{-3}$  & 202 & 201 & 187 & 182 & 179 \\
\hline\hline
$e^+ e^- \rightarrow W^+W^- $  & $0.12714$ & 19010 & 2331 & 266 & 151 & 127 \\
\hline
$e^+ e^- \rightarrow\tau^+\tau^- $  &$1.562 \times 10^{-6}$  & 1 & - & - & - & - \\
\hline
$e^+ e^- \rightarrow t\bar t $  &$0.0153$  & 29 & 18 & 5 & 3 & 2 \\
\hline
$e^+ e^- \rightarrow ZZ $  & $3.188 \times 10^{-3}$ & 339 & 15 & 7 & 3 & 3 \\
\hline
$e^+ e^- \rightarrow Zh $  & $1.533 \times 10^{-4}$ & 4 & 4 & 4 & 2 & 2 \\
\hline
$e^+ e^- \rightarrow W^+W^-Z $  & $9.814 \times 10^{-4}$ & 154 & 75 & 53 & 23 & 21 \\
\hline
$e^+ e^- \rightarrow ZZZ $  &$9.51 \times 10^{-6}$  & 1 & 1 & 1 & 1 & 1 \\
\hline
$e^+ e^- \rightarrow t\bar tZ $  & $4.64 \times 10^{-3}$ & - &- &- & - &-\\
\hline
$e^+ e^- \rightarrow ZZh $  & $3.21 \times 10^{-4}$ & - &- &- &- &-  \\
\hline
$e^+ e^- \rightarrow \ell jj\nu $  & $1.166$ & 1 & 1 & 1 & 1 &- \\
\hline
\hline
\end{tabular}
\caption{Cross-sections of the signal and various background channels for leptonic decay of $\tau$, are shown in pb alongside the number of expected events for the
individual channels at 500 $\ifb$ luminosity after applying each of the cuts {\bf C0} - {\bf C4} as listed in the text. NEV $\equiv$ number of events. 
All the numbers are presented for BR$(h \rightarrow \mu \tau) = 9.78 \times 10^{-3}$.}
\label{tab:vvh_2l_1000}
\end{center}
\end{table}%
%%%%%%%%%%%%%%%%%%%%%%%%%%%%%%%%%%%%%%%%%%%%%%%%%%%%%

As evident from the numbers in table~\ref{tab:vvh_2l_1000}, $WW$ production channel is the most dominant contributor to the SM background. 
The cuts {\bf C1}, {\bf C2} and {\bf C3} are particularly effective in reducing this background. Besides, {\bf C2} also reasonably 
reduces the two other potentially dominant channels, $ZZ$ and $WWZ$. {\bf C1} and {\bf C2} are helpful in reducing the $t\bar t$ background. 
Overall, one can achieve a 3$\sigma$ statistical significance at $\mathcal{L} \approx 30~\ifb$ which is a large improvement 
over the $\sqrt{s}=250$ GeV analysis.
%%%%%%%%%%%%%%%%%%%%%%%%%%
%%%%%%%%%%%%%%%%%%%%%%%%%%
\item {\bf Final state: $1\mu + 1\tau{\rm -jet} + \me$ :} \\
%\label{sec:hvv_had}
%%%%%%%%%%%%%%%%%%%%%%%%%%%%%%%%%%%%%%%%%%%%%%%%%%%%%
For the final state $1\mu + 1\tau{\rm -jet} + \me$ we have used the following kinematical cuts:
\begin{itemize}
\item {\bf D0 :} In the final state, we demand one muon along with a jet which must be tagged as a $\tau$-jet. Any 
additional leptons and jets in the event including $b$-jets have been vetoed.
\item {\bf D1 :} The missing energy distribution is expected to be slightly on the softer side than that in the
$\tau$ leptonic decay case. The normalized distribution of $\me$ have been shown in Fig. \ref{fig:me_had_1000} for the signal as well as the 
same SM background channels with similar color coding as in Fig.~\ref{fig:me_lep_1000}. We demand $1000~{\rm GeV} > \me > 500~{\rm GeV}$.
%%%%%%%%%%%%%%%%%%%%%%%%%%%%
\begin{figure}[h!]
\begin{center}
\includegraphics[scale=0.45]{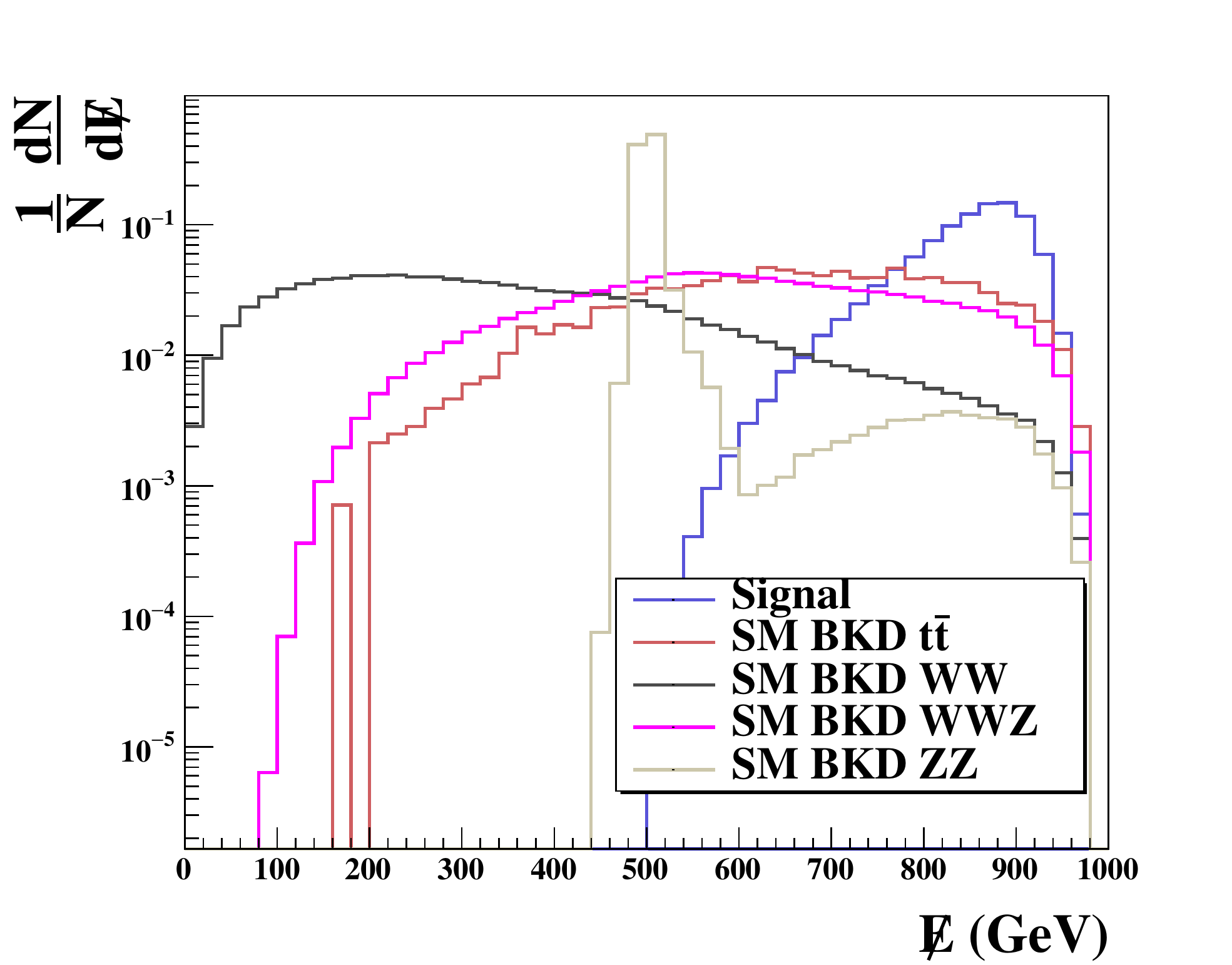}
\caption{Normalized $\me$ distribution for signal and backgrounds at $\sqrt{s}=1000$ GeV for final state: $1\mu + 1\tau{\rm -jet} + \me$.}
\label{fig:me_had_1000}
\end{center}
\end{figure}
%%%%%%%%%%%%%%%%%%%%%%%%%%%%
\item {\bf D2 :} We demand that the visible invariant mass, that is the visible mass of the muon and $\tau$-jet system
should lie within the region $130~{\rm GeV} > M_{\mu\tau_{\rm had}} > 70~{\rm GeV}$ following the distribution in Fig.\ref{M_mutau_had}.
%%%%%%%%%%%%%%%%%%%%%%%%%%%%
\begin{figure}[h!]
\begin{center}
\includegraphics[scale=0.45]{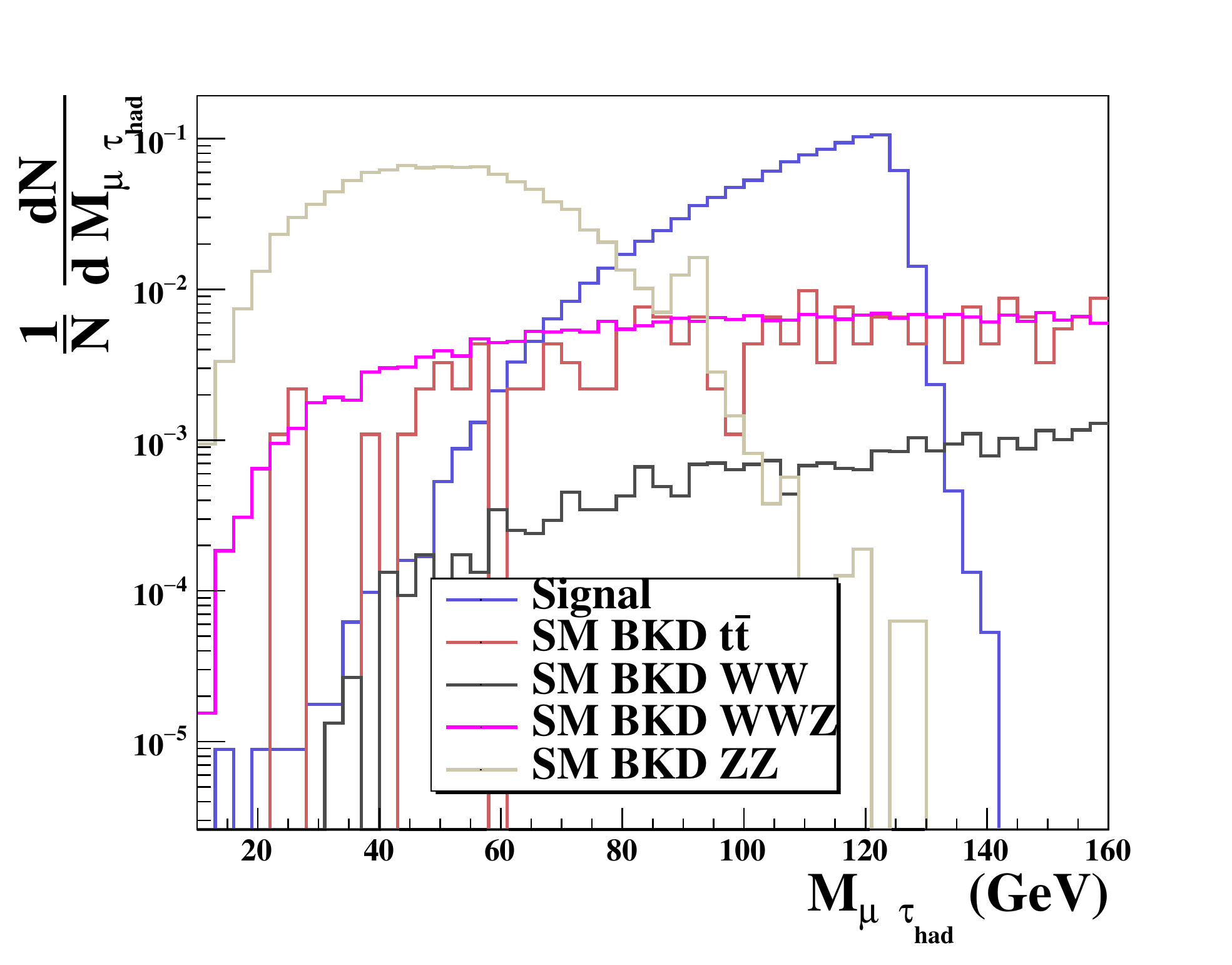}
\caption{Normalized $M_{\mu\tau_{\rm had}}$ distribution for signal and backgrounds at $\sqrt{s}=1000$ GeV for final state: $1\mu + 1\tau{\rm -jet} + \me$.}
\label{M_mutau_had}
\end{center}
\end{figure}
%%%%%%%%%%%%%%%%%%%%%%%%%%%%
\item {\bf D3 :} We demand that the visible momentum, that is the visible momentum of the muon and $\tau$-jet system
should lie within the region $320~{\rm GeV} > p^{vis} > 20~{\rm GeV}$. Corresponding distribution is shown in Fig.~\ref{p_vis_had}.
%%%%%%%%%%%%%%%%%%%%%%%%%%%%
\begin{figure}[h!]
\begin{center}
\includegraphics[scale=0.45]{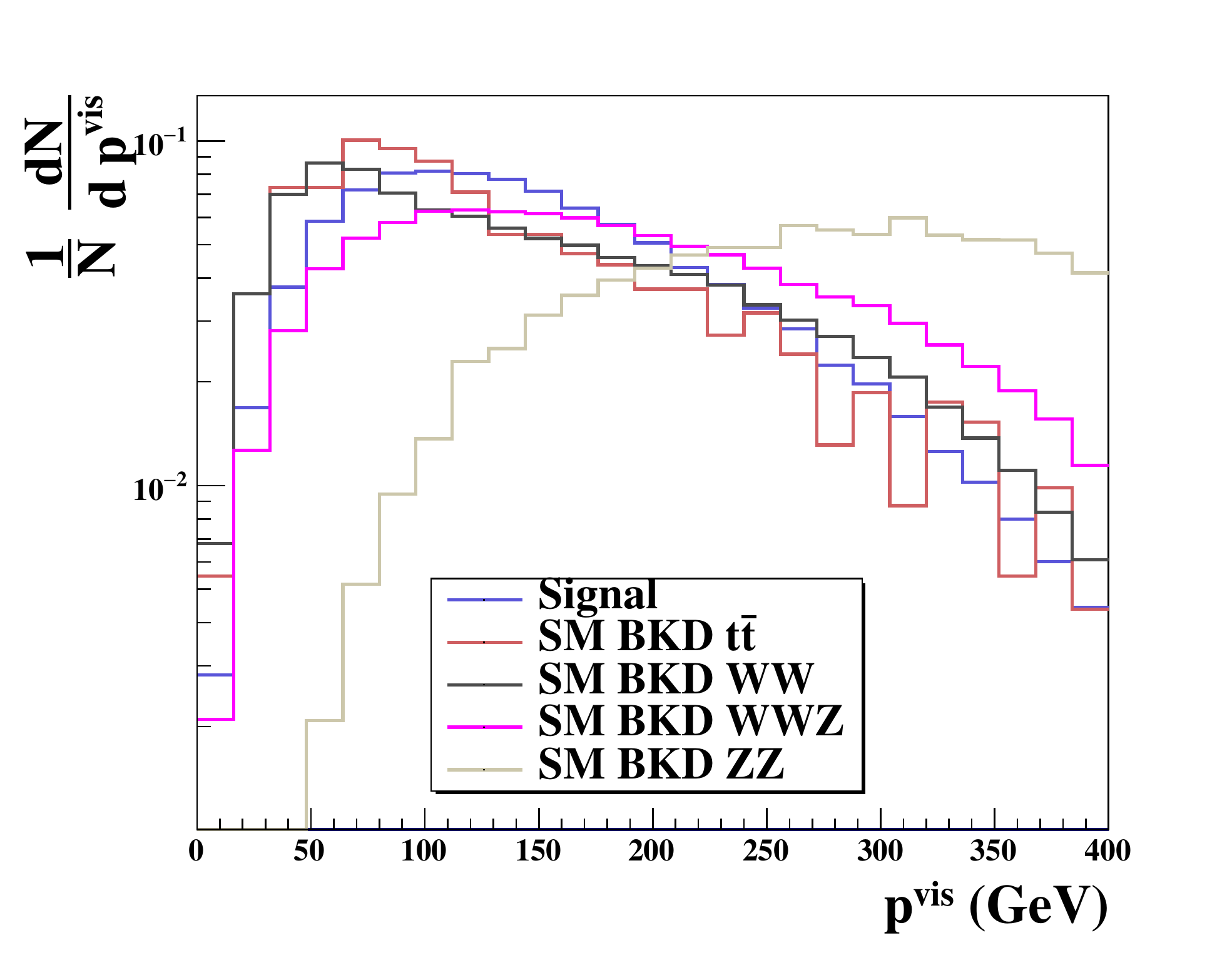}
\caption{Normalized $ p^{vis}$ distribution for signal and backgrounds at $\sqrt{s}=1000$ GeV for final state: $1\mu + 1\tau{\rm -jet} + \me$.}
\label{p_vis_had}
\end{center}
\end{figure}
%%%%%%%%%%%%%%%%%%%%%%%%%%%%
\item {\bf D4 :} In our signal events, we expect the missing energy vector, $\vec{\me}$ to be well separated from this $\tau$-jet. We demand,
$5.5 > \Delta R(\tau{\rm -jet},\vec{\me}) > 1.5$. The normalized distribution of $\Delta R(\tau{\rm -jet},\vec{\me})$ is shown in Fig.~\ref{del_R_1000}.
%%%%%%%%%%%%%%%%%%%%%%%%%%%%
\begin{figure}[h!]
\begin{center}
\includegraphics[scale=0.45]{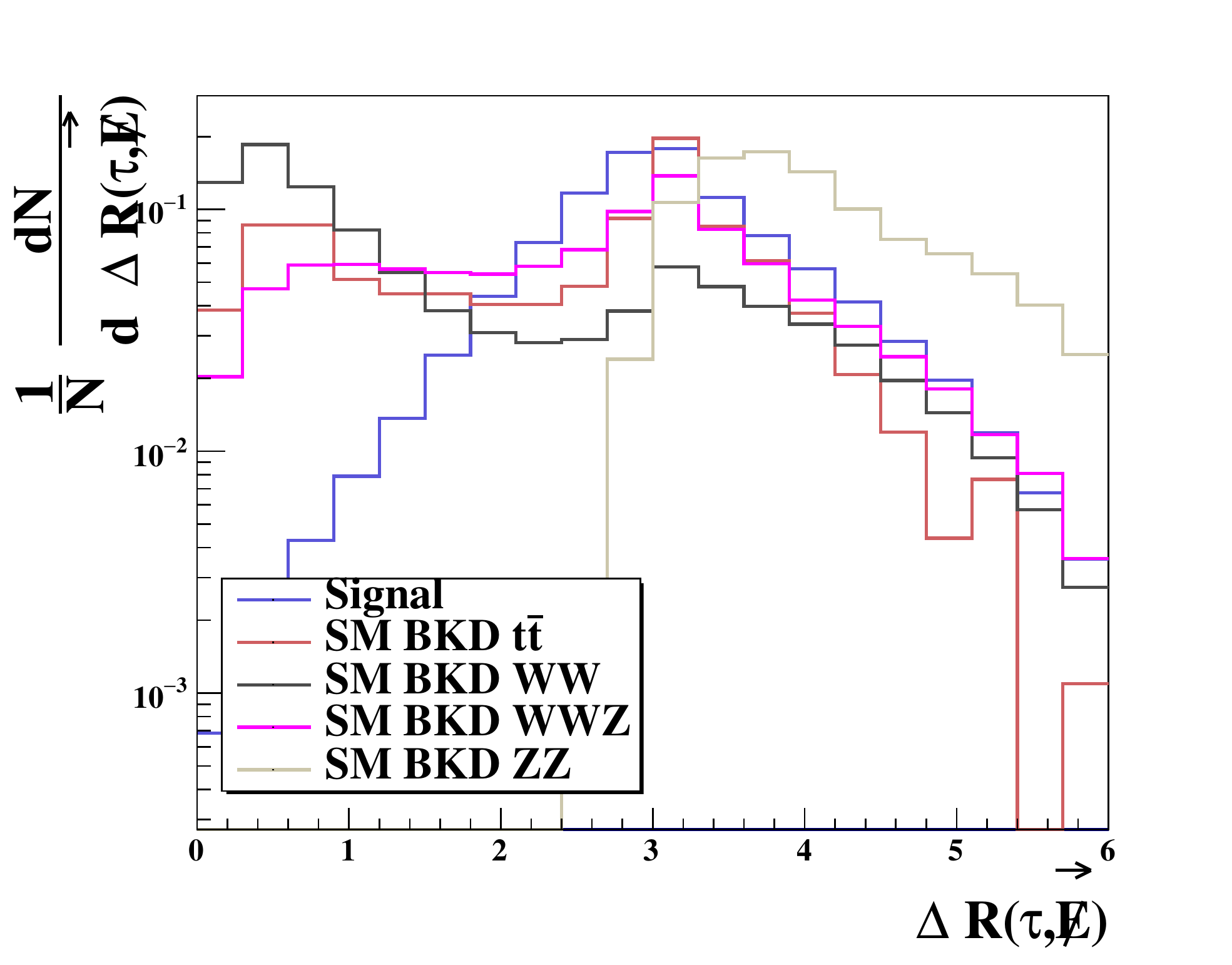}
\caption{Normalized $ \Delta R(\tau{\rm -jet},\vec{\me})$ distribution for signal and backgrounds at $\sqrt{s}=1000$ GeV for final state: 
$1\mu + 1\tau{\rm -jet} + \me$.}
\label{del_R_1000}
\end{center}
\end{figure}
%%%%%%%%%%%%%%%%%%%%%%%%%%%%
\end{itemize}
In Table~\ref{tab:vvh_1l1t_1000} below we have presented the cut-flow numbers obtained from our collider simulation at 
$\sqrt{s}=1000$ GeV and an integrated luminosity of ${\mathcal{L}}=500~\ifb$ corresponding to our signal $1\mu + 1\tau{\rm -jet} + \me$ (with 
$BR(h \rightarrow \mu \tau) = 9.78 \times 10^{-3}$) as well as the different SM background channels.  
%%%%%%%%%%%%%%%%%%%%%%%%%%%%%%%%%%%%%%%%%%%%%%%%%%%%n{table}[ht!]
%%\scriptsize
\begin{table}[h]
\begin{center}
\begin{tabular}{||c|c|c|c|c|c|c||}
\hline
\multicolumn{1}{||c|}{\bf Process} &
\multicolumn{6}{|c||}{\bf $\sqrt{s}=$1000 GeV} \\
\cline{2-7}
\multicolumn{1}{||c|}{} &
\multicolumn{1}{|c|}{$\sigma$ (pb)} &
\multicolumn{5}{|c||}{NEV ($\mathcal{L}=$500 $\ifb$)} \\
\cline{3-7}
 &  & {\bf D0} & {\bf D1} & {\bf D2} & {\bf D3} & {\bf D4} \\
\hline
$e^+ e^- \rightarrow \nu_e\bar{\nu}_e h$  &$2.01 \times 10^{-3}$   & 226 & 226 & 221 & 209 & 201 \\
\hline\hline
$e^+ e^- \rightarrow W^+W^- $  & $0.12714$ & 4778 & 1413 & 59 & 56 & 48 \\
\hline
$e^+ e^- \rightarrow\tau^+\tau^- $  &$1.562 \times 10^{-6}$  & - & - & - &- &- \\
\hline
$e^+ e^- \rightarrow t\bar t $  &$0.0153$  & 9 & 7 & 1 & 1 & 1\\
\hline
$e^+ e^- \rightarrow ZZ $  & $3.188 \times 10^{-3}$ & 25 & 25 & 3 & -& -\\
\hline
$e^+ e^- \rightarrow Zh $  & $1.533 \times 10^{-4}$ & 7 & 6& 2 & 1 & 1 \\
\hline
$e^+ e^- \rightarrow W^+W^-Z $  & $9.814 \times 10^{-4}$ & 32 & 24 & 4 & 3 & 3\\
\hline
$e^+ e^- \rightarrow ZZZ $  &$9.51 \times 10^{-6}$  & - & - & - &- &- \\
\hline
$e^+ e^- \rightarrow t\bar tZ $  & $4.64 \times 10^{-3}$ & - &- & - &- &-\\
\hline
$e^+ e^- \rightarrow ZZh $  & $3.21 \times 10^{-4}$ & - &- &- &- &-  \\
\hline
$e^+ e^- \rightarrow \ell jj\nu $  & $1.166$ & 634 & 113 & 1 & 1 &1 \\
\hline
\hline
\end{tabular}
\caption{Cross-sections of the signal and various background channels for hadronic decay of $\tau$, are shown in pb alongside the number of expected events for the
individual channels at 500 $\ifb$ luminosity after applying each of the cuts {\bf D0} - {\bf D4} as listed in the text. NEV $\equiv$ number of events. 
All the numbers are presented for BR$(h \rightarrow \mu \tau) = 9.78 \times 10^{-3}$.}
\label{tab:vvh_1l1t_1000}
\end{center}
\end{table}%
%%%%%%%%%%%%%%%%%%%%%%%%%%%%%%%%%%%%%%%%%%%%%%%%%%%%%
\end{itemize}
It is evident from Table \ref{tab:vvh_1l1t_1000} that the dominant SM backgrounds are
$W^+ W^-$, $ZZ$ and $W^+ W^- Z$. However, these contributions are effectively reduced by the cut {\bf D1} and then
gradually cut down by the  [{\bf D2} - {\bf D4}]. It is worth noting that for the $e^+e^-\to\nu_e\bar\nu_e h$ production mode, 
we have used {\bf C3} (for leptonic $\tau$-decay) and {\bf D2} (for hadronic $\tau$-decay) which restrict the visible invariant 
mass of the two lepton system and $\mu$-$\tau$-jet system respectively and not on the collinear mass, as used for $e^+e^-\to Zh$ 
production mode. This is because,  the collinear mass cannot be constructed
whenever there are additional source(s) of missing energy over and above $\tau$-decay. As the numbers in Table~\ref{tab:vvh_1l1t_1000} indicate, a 3$\sigma$ statistical significance may be obtained 
at a very low integrated luminosity of $12~\ifb$. This still is a slight improvement over what is obtained for the $2\ell + \me$ final state.  

Hence the $1\mu + 1\tau{\rm -jet} + \me$ final state at $\sqrt{s}=1000$ GeV has the potential to probe the smallest  
BR$(h \rightarrow \mu \tau) (\sim 10^{-4})$  than all other final states studied so far.  
The lowest possible branching ratios that can be probed at 3$\sigma$ statistical significance with the two final states studied 
at this center-of-mass energy have been shown at three different integrated luminosities in Table~ \ref{lowest_br_aw_1000}.
%%%%%%%%%%%%%%%%%%%%%%%%%%%%%%%%%%%%%%%%%%%%
\begin{table}
\begin{center}
\begin{tabular}{|| c | c | c | c ||}
\hline
$\mathcal{L} (fb^{-1})$ & BR in $(2 \ell +\cancel{E})$ & BR in $(\mu +\tau_{had}+\cancel{E})$ & Combined BR\\ \hline
250 & $3.16\times 10^{-3}$ & $1.58\times 10^{-3}$ & $1.62 \times 10^{-3}$ \\
500 & $2.15 \times 10^{-3}$ & $1.11\times 10^{-3}$ & $1.11 \times 10^{-3}$\\
1000 & $1.44 \times 10^{-3}$ & $7.22\times 10^{-4}$ & $7.58 \times 10^{-4}$ \\
\hline
\end{tabular}
\caption{Lowest branching ratio that can be probed with 3$\sigma$ statistical significance for the two different final states (arising from leptonic and hadronic decay of $\tau$) at $\sqrt{s}$=1000 GeV with BR$(h \rightarrow \mu \tau) = 9.78 \times 10^{-3}$. The last column indicates the BR reach when the event rates of these two final states are combined together.}
\label{lowest_br_aw_1000}
\end{center}
\end{table}

Note that, the collider analyses presented so far at two different 
center-of-mass energies have been performed for specific choices of $a_Z$ and $a_W$. Although the allowed ranges of these 
parameters are quite constrained as discussed in section~\ref{sec:frame_hvv_mult}, it would be interesting to see 
how the collider reach in terms of the relevant branching ratio varies along their whole allowed ranges. We have depicted 
this below in Fig.~\ref{fig:aw_az_reach}.   
%%%%%%%%%%%%%%%%%%%%%%%%%%%%%%
\begin{figure}[h!]
\begin{center}
\includegraphics[scale=0.30]{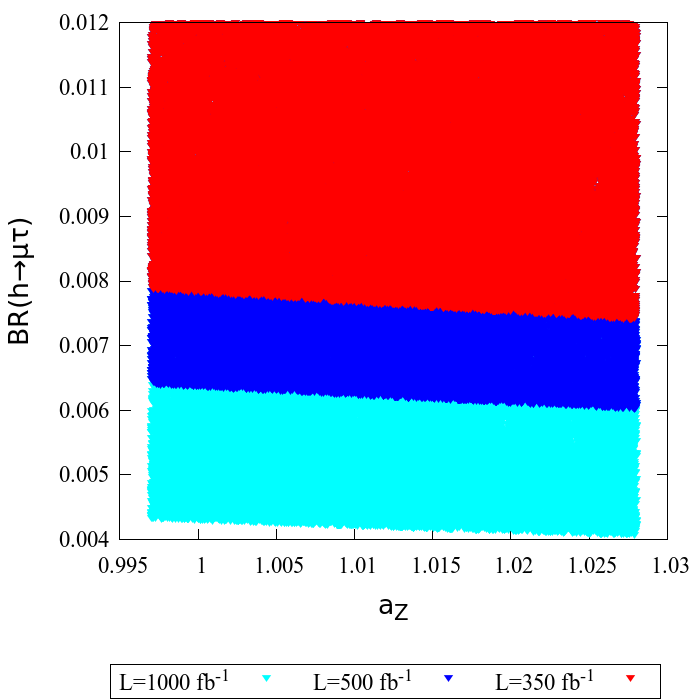}
\includegraphics[scale=0.30]{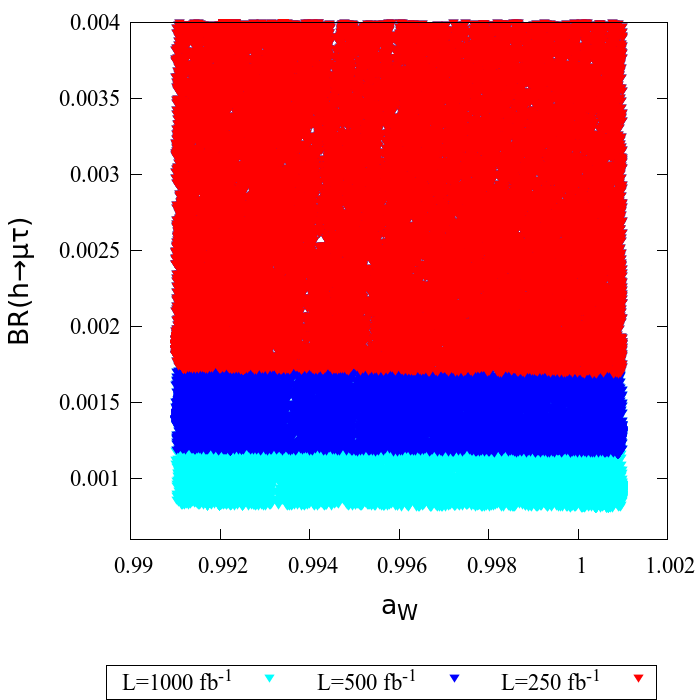}
\caption{The $3\sigma$ reach of BR$(h\to\mu\tau)$ at $\sqrt{s}=250$ GeV and 1000 GeV over the allowed ranges of $a_Z$ and $a_W$ respectively 
at different integrated luminosities.}
\label{fig:aw_az_reach}
\end{center}
\end{figure}
%%%%%%%%%%%%%%%%%%%%%%%%%%%%%%

The red color indicates $3\sigma$ reach of BR$(h\to\mu\tau)$ at $350~\ifb$ and $250~\ifb$ luminosities at $\sqrt{s}=250$ GeV and 1000 GeV 
respectively. Similarly, the blue and cyan colors indicate the reach of the same at $500~\ifb$ and $1000~\ifb$ luminosities at both 
the center-of-mass energies. As evident from the plots, the branching ratio does not vary much so as to make any visible changes in the 
predicted results over the presently allowed regions of $a_Z$ and $a_W$. 
%%%%%%%%%%%%%%%%%%%%%%%%%%%%%%%%
\subsection{Prospects of higher-dimensional operators}
\label{sec:coll_hdm}
%%%%%%%%%%%%%%%%%%%%%%%%%%%%%%%%
As discussed earlier, introducing effective operators may enhance the prospects of probing even smaller BR($h \rightarrow \mu \tau$) 
by enhancing the production cross-section of the Higgs boson due to their momentum-dependent Lorentz structures. 
From Table \ref{table1} it can be seen that all the four non-zero $f_n$'s, {\em i.e.} $f_W$ , $f_{B}$, $f_{WW}$, $f_{BB}$ 
can modify the $hZZ$ interaction ( $g_{h Z Z}^{(1)}$ , $g_{h Z Z}^{(2)}$). On the other hand, $f_W$ and $f_{WW}$ can modify 
the $hWW$ interaction. Since the sole purpose of introducing these operators is to assess whether they can improve the reach on 
smaller BR($h \rightarrow \mu \tau$), we first proceed to study how much enhancement in the Higgs boson production cross-section 
one can expect from the presence of these operators. In order to determine that, we have used conservative values of $f_n$'s for our analysis, compared to their maximally allowed values as mentioned in Table~\ref{table0}. Non-zero values of $f_n$'s result in enhancement of the Higgs production cross-section and allow us to probe even smaller BR($h \rightarrow \mu \tau$). Higgs production cross-sections for some sample values of $f_n$'s are given in Table~\ref{tab:cs_hdo}. Less conservative, 2$\sigma$ allowed values of $f_n$'s as mentioned in Table~\ref{table0}, would thus indeed improve the reach of $e^+ e^-$ collider in probing the lowest possible branching ratio.  
%%%%%%%%%%%%%%%%%%%%%%%%%%%%%%%%%%%%%%
\begin{table}[h!]
\begin{center}
\begin{tabular}{ | c | c| c | c |}
\hline
 Couplings & Values of the couplings &$\sigma_{prod}$ at $\sqrt{s} = 250$~GeV in pb &  $\sigma_{prod}$ at $\sqrt{s} = 1000$~GeV in pb \\ \hline 
 $f_B$ & -3.4 & 0.2514 & -  \\ 
 & 11.0 & 0.2329 & - \\ \hline 
 $f_{BB}$  &  -2.78 & 0.2503 & - \\ 
 & 0.283 & 0.2466 & - \\ \hline
 $f_W$ &  -5.8 & 0.2737 & 0.1959\\
 & 14.5 & 0.187 & 0.3059 \\ \hline
 $f_{WW}$ & -1.86 & 0.2814 & 0.2256 \\
  & 0.5 & 0.2463 & 0.2284 \\
 \hline 
\end{tabular}
\caption{Production cross-sections ($\sigma_{prod}$) of Higgs boson at $\sqrt{s} = 250$  GeV and $\sqrt{s} = 1000$ 
GeV for some sample values of $f_n$ with $\Lambda = 1$ TeV.}
\label{tab:cs_hdo}
\end{center}
\end{table}
%%%%%%%%%%%%%%%%%%%%%%%%%%%%%%%%%%%%%%%%%

As can be seen from Table \ref{tab:cs_hdo}, an enhancement in $e^+e^-\to Zh$ production cross-section at $\sqrt{s}=250$ GeV is obtained 
for a sample value $f_{WW} = -1.86$ (value of $f_{WW}$ being compatible with electroweak precision observables and signal strengths mentioned earlier), 
while keeping $f_W$, $f_B$ and $f_{BB}$ zero. For the sake of improvement of results, we have narrowed down the collinear mass cut {\bf {[A3, B3]}} (mentioned earlier) a little, and varied 
$ M_{\rm coll}$ as ,$(m_h + 12)~{\rm GeV} > M_{\rm coll} > (m_h - 12)~{\rm GeV}$. However, the enhancement can be at most by a factor $\approx 1.10$ which is not enough 
to increase the signal significance sufficiently so as to improve upon our results obtained for $\sqrt{s}=1000$ GeV analysis.
\footnote{Our analysis reveals that the combination of the two final states at $\sqrt{s}=250$ GeV with an integrated luminosity of 
$1000~\ifb$ results in a reach of ${\rm BR}(h \rightarrow \mu \tau)\approx 2.69 \times 10^{-3}$  which is barely smaller 
by a factor of $\approx 2$ compared to that obtained in the absence of $f_{WW}$.} Similarly for the $e^+e^-\to\nu_e\bar\nu_e h$ production channel, 
an enhancement in the cross-section is obtained for the sample value  $f_W=14.5$ keeping all the other $f_n$'s zero at $\sqrt{s}=1000$ GeV.
We have subsequently carried a detailed simulation for this case. The results are presented in Table \ref{tab:vvh_2l_1000_fw} and 
\ref{tab:vvh_1l1t_1000_fw} for the final states $2\ell + \me$ and $1\mu + 1\tau{\rm -jet} + \me$ respectively. 

The signal and backgrounds will remain same as before. At $\sqrt{s} = 1000$ GeV, signal cross-section increases from $2.01 \times 10^{-3}$ pb (earlier scenario) 
to $2.691 \times 10^{-3}$ pb and all the background cross-sections except $W^+ W^- Z$ remain unaltered as can be seen from 
Table~ \ref{tab:vvh_2l_1000_fw} and Table~ \ref{tab:vvh_1l1t_1000_fw}. The numbers are presented for $\mathcal{L} = 500~\ifb$ and 
BR$(h \rightarrow \mu \tau) = 9.78 \times 10^{-3}$ as before.
%%%%%%%%%%%%%%%%%%%%%%%%%%%%%%%%%%%%%%%%%%%%%%
\begin{table}[h]
\begin{center}
\begin{tabular}{||c|c|c|c|c|c|c||}
\hline
\multicolumn{1}{||c|}{\bf Process} &
\multicolumn{6}{|c||}{\bf $\sqrt{s}=$1000 GeV} \\
\multicolumn{1}{||c|}{} &
\multicolumn{6}{|c||}{(with HDO)} \\ 
\cline{2-7}
\multicolumn{1}{||c|}{} &
\multicolumn{1}{|c|}{$\sigma$ (pb)} &
\multicolumn{5}{|c||}{NEV ($\mathcal{L}=$500 $\ifb$)} \\
\cline{3-7}
 &  & {\bf C0} & {\bf C1} & {\bf C2} & {\bf C3} & {\bf C4}  \\
\hline
$e^+ e^- \rightarrow \nu_e\bar{\nu}_e h$  &$2.691 \times 10^{-3}$ & 276 & 274 & 258 & 250 & 245  \\
\hline\hline
$e^+ e^- \rightarrow W^+W^- $  & $0.12714$ & 19010 & 2331 & 266 & 151 & 127 \\
\hline
$e^+ e^- \rightarrow\tau^+\tau^- $  &$1.562 \times 10^{-6}$  & 1 & - & - & - & - \\
\hline
$e^+ e^- \rightarrow t\bar t $  &$0.0153$  & 29 & 18 & 5 & 3 & 2 \\
\hline
$e^+ e^- \rightarrow ZZ $  & $3.188 \times 10^{-3}$ & 339 & 15 & 7 & 3 & 3 \\
\hline
$e^+ e^- \rightarrow Zh $  & $1.533 \times 10^{-4}$ & 4 & 4 & 4 & 2 & 2 \\
\hline
$e^+ e^- \rightarrow W^+W^-Z $  & $2.38\times 10^{-4}$ & 39 & 33 & 28 & 10 & 10 \\
\hline
$e^+ e^- \rightarrow ZZZ $  &$9.51 \times 10^{-6}$  & 1 & 1 & 1 & 1 & 1 \\
\hline
$e^+ e^- \rightarrow t\bar tZ $  & $4.64 \times 10^{-3}$ & - &- &- & - &-\\
\hline
$e^+ e^- \rightarrow ZZh $  & $3.21 \times 10^{-4}$ & - &- &- &- &-  \\
\hline
$e^+ e^- \rightarrow \ell jj\nu $  & $1.166$ & 1 & 1 & 1 & 1 &- \\
\hline
\hline
\end{tabular}
\caption{Cross-sections of the signal and various background channels for leptonic decay of $\tau$, are shown in pb alongside the number of 
expected events for the individual channels at 500 $\ifb$ luminosity after applying each of the cuts {\bf C0} - {\bf C4} 
as listed in the text. NEV $\equiv$ number of events. All the numbers are presented for 
$f_{WW} = 14.0$, BR$(h \rightarrow \mu \tau) = 9.78 \times 10^{-3}$.}
\label{tab:vvh_2l_1000_fw}
\end{center}
\end{table}%
%%%%%%%%%%%%%%%%%%%%%%%%%%%%%%%%%%%%%%%%%%%%%%
\begin{table}[h]
\begin{center}
\begin{tabular}{||c|c|c|c|c|c|c||}
\hline
\multicolumn{1}{||c|}{\bf Process} &
\multicolumn{6}{|c||}{\bf $\sqrt{s}=$1000 GeV} \\
\multicolumn{1}{||c|}{} &
\multicolumn{6}{|c||}{(with HDO)} \\ 
\cline{2-7}
\multicolumn{1}{||c|}{} &
\multicolumn{1}{|c|}{$\sigma$ (pb)} &
\multicolumn{5}{|c||}{NEV ($\mathcal{L}=$500 $\ifb$)} \\
\cline{3-7}
 &  & {\bf D0} & {\bf D1} & {\bf D2} & {\bf D3} & {\bf D4} \\
\hline
$e^+ e^- \rightarrow \nu_e\bar{\nu}_e h$  &$2.691  \times 10^{-3}$   & 315 & 315 & 305 & 276 & 267 \\
\hline\hline
$e^+ e^- \rightarrow W^+W^- $  & $0.12714$ & 4778 & 1413 & 59 & 56 & 48 \\
\hline
$e^+ e^- \rightarrow\tau^+\tau^- $  &$1.562 \times 10^{-6}$  & - & - & - &- &- \\
\hline
$e^+ e^- \rightarrow t\bar t $  &$0.0153$  & 9 & 7 & 1 & 1 & 1\\
\hline
$e^+ e^- \rightarrow ZZ $  & $3.188 \times 10^{-3}$ & 25 & 25 & 3 & -& -\\
\hline
$e^+ e^- \rightarrow Zh $  & $1.533 \times 10^{-4}$ & 7 & 6& 2 & 1 & 1 \\
\hline
$e^+ e^- \rightarrow W^+W^-Z $  & $2.38\times 10^{-4}$ & 8 & 7 & 2 & 2 & 2\\
\hline
$e^+ e^- \rightarrow ZZZ $  &$9.51 \times 10^{-6}$  & - & - & - &- &- \\
\hline
$e^+ e^- \rightarrow t\bar tZ $  & $4.64 \times 10^{-3}$ & - &- & - &- &-\\
\hline
$e^+ e^- \rightarrow ZZh $  & $3.21 \times 10^{-4}$ & - &- &- &- &-  \\
\hline
$e^+ e^- \rightarrow \ell jj\nu $  & $1.166$ & 634 & 113 & 1 & 1 &1 \\
\hline
\hline
\end{tabular}
\caption{Cross-sections of the signal and various background channels for hadronic decay of $\tau$, are shown in pb alongside the number of 
expected events for the individual channels at 500 $\ifb$ luminosity after applying each of the cuts {\bf D0} - {\bf D4} 
as listed in the text. NEV $\equiv$ number of events. All the numbers are presented for 
$f_{WW} = 14.0$ , BR$(h \rightarrow \mu \tau) = 9.78 \times 10^{-3}$.}
\label{tab:vvh_1l1t_1000_fw}
\end{center}
\end{table}%
%%%%%%%%%%%%%%%%%%%%%%%%%%%%%%%%%%%%%%%%%%%%%%%%%%%%%%%%%%%%%

Table~\ref{lowest_br_1000_fW} shows slight improvement in probing BR($h \rightarrow \mu \tau$).
The combined result from the two channels gives the best reach of branching ratio ($\approx 5.83\times 10^{-4}$) which is 
an improvement by a factor of $\approx 1.24$ over that obtained in absence of $f_{WW}$ and it
 is the best reach obtained at $e^+ e^-$ collider at 1000 GeV.
%%%%%%%%%%%%%%%%%%%%%%%%%%%%%%%%%%%%%%%%%%%%%%%%%%%%%%%%%%%%%
\begin{table}
\begin{center}
\begin{tabular}{|| c | c | c | c ||}
\hline
$\mathcal{L} (fb^{-1})$ & BR in $(2 \ell +\cancel{E})$ & BR in $(\mu +\tau_{had}+\cancel{E})$ & Combined BR\\ \hline
250 & $2.15\times 10^{-3}$ & $1.23\times 10^{-3}$ & $1.19 \times 10^{-3}$ \\
500 & $1.49\times 10^{-3}$ & $8.51\times 10^{-4}$ & $8.33 \times 10^{-4}$\\
1000 & $1.05 \times 10^{-3}$ & $5.91\times 10^{-4}$ & $5.83 \times 10^{-4}$ \\
\hline
\end{tabular}
\caption{Lowest branching ratio that can be probed with 3$\sigma$ statistical significance for the two different final states (arising from leptonic and hadronic decay of $\tau$) at $\sqrt{s}$=1000 GeV with 
$f_{WW} = 14.0$ and BR$(h \rightarrow \mu \tau) = 9.78 \times 10^{-3}$. The last column indicates the BR reach when the event rates of these two final states are combined together. }
\label{lowest_br_1000_fW}
\end{center}
\end{table}

It can therefore be concluded that at $\sqrt{s} = 1000$ GeV and $\mathcal{L} = 1000$ fb$^{-1}$, $e^+ e^-$ collider provides at least two orders of magnitude improvement in probing $h \rightarrow \mu \tau$ branching ratio as compared to  the existing limits at LHC. It is because of its relatively clean environment. At $\sqrt{s}=1000$ GeV, the number of signals surviving is much larger than the number of total backgrounds after applying all the cuts. This enhances the signal significance and $3\sigma$ significance is achieved at very low luminosity for a the fixed value of BR$(h \rightarrow \mu \tau) = 9.78 \times 10^{-3}$. Thus the branching ratio as small as $\sim 10^{-4}$ can be probed by enhancing the integrated luminosity.
%%%%%%%%%%%%%%%%%%%%%%%%%%%%%%%%%%%%%%%%%%%%%%%%%%%%%%%%%%%%%
\section{Summary and Conclusions}
\label{sec:concl}
%%%%%%%%%%%%%%%%%%%%%%%%%%%%%%%%%%%%%%%%%%%%%%%%%%%%%%%%%%%%%
The objective of this work was to study the collider aspects of one of these possible non-standard 
decay modes, namely, $h \rightarrow \mu \tau$ and examine the possible reach of the corresponding branching ratio at future $e^+ e^-$ colliders. 
Collider simulation has been performed at $\sqrt{s} = $250 GeV and 1000 GeV at 
three projected integrated luminosities, {\em i.e.} $\mathcal{L}= $ 350 (250) fb$^{-1}$, 500 fb$^{-1}$, 1000 fb$^{-1}$. 
We have explored different possible final states arising from both leptonic and hadronic decays of the $\tau$. We have looked for the smallest possible BR($h\to\mu\tau$) that can be probed at the 3$\sigma$ level. We have also combined the event rates of different possible final states at same centre-of-mass energy to improve the reach. Two different scenarios have been considered separately for this purpose, with two different types of modifications at the production level of Higgs boson. The first scenario includes modification of $hVV$ interaction with multiplicative factors only (achieved by scaling the vertex factor), whereas effective operators with new Lorentz structures have been introduced in the second scenario. While introducing the effective operators, we have chosen the effective couplings ($f_n$) in a somewhat conservative manner, though the production cross-section of Higgs boson gets enhanced. In principle, one can also use the values of $f_n$'s (allowed by the 2$\sigma$ constraints), which could lead to larger production cross-section and would be useful in probing even lower branching ratios.

At $\sqrt{s} = 250$ GeV, $e^+e^-\to Zh$ is the main production mode of the Higgs boson. The lowest branching ratio that can be probed at $3 \sigma$ level is $\approx 4.09\times 10^{-3}$ at an integrated luminosity, $\mathcal{L} = 1000~\ifb$. The result improves  slightly after including the effective operators instead of simply scaling the $hVV$ vertices, though the order of magnitude of the lowest detectable branching ratio remains the same.

At $\sqrt{s} = 1000$ GeV, the reach of BR$(h \rightarrow \mu \tau)$ is much better owing to the large Higgs production cross-section in the $e^+e^-\to h\nu_e\bar\nu_e$ mode. Combining the signal rates in the two aforementioned final states at this centre-of-mass energy, one can probe BR($h \rightarrow \mu \tau$) down to  $\approx 5.83\times 10^{-4}$ with a $3 \sigma$ statistical significance 
at $\mathcal{L} = 1000~\ifb$. This is the best reach so far, which an $e^+ e^-$ collider can achieve, and is smaller by nearly two orders of magnitude than what is obtained from the latest LHC data. 
%%%%%%%%%%%%%%%%%%%%%%%%%%%%%%%%%%%%%%%%%%%%%%%%%%%%%%%%%%%%%%%%%%%%%%%%%%%%%%%%%%%%%%%%%%%%%%%%%%%%%%%%%%%%%%%%%%%%%%%%%%%%%%%%%%%%%%%%%%%%%%%%%%%
\section{Acknowledgement}
%%%%%%%%%%%%%%%%%%%%%%%%%%%%%
We thank Nabarun Chakrabarty, Shankha Banerjee and Biplob Bhattacharjee for fruitful discussions.
This work is partially supported by funding available from the Department of Atomic Energy, Government of India, 
for the Regional Center for Accelerator- based Particle Physics (RECAPP), Harish-Chandra Research Institute. 
Computational work for this study was carried out at the cluster computing facility in the Harish-Chandra Research 
Institute (http://www.hri.res.in/cluster).
%%%%%%%%%%%%%%
\bibliography{hmutau_ilc}{}
%%%%%%%%%%%%%%
\end{document}